# An Improved RIP-Based Performance Guarantee for Sparse Signal Recovery via Orthogonal Matching Pursuit

Ling-Hua Chang and Jwo-Yuh Wu

Department of Electrical and Computer Engineering,
Institute of Communications Engineering,
National Chiao Tung University,
1001, Ta-Hsueh Road, Hsinchu, Taiwan
Tel: 886-35-712121, ext. 54524; Fax: 886-35-710116
Email: iamjaung@gmail.com, jywu@cc.nctu.edu.tw

*Abstract* - **A sufficient condition reported very recently for perfect recovery of a *K*-sparse vector via orthogonal matching pursuit (OMP) in *K* iterations is that the restricted isometry constant of the sensing matrix satisfies** $\delta_{K+1} < \frac{1}{\sqrt{K}+1}$. **By exploiting an approximate orthogonality condition characterized via the achievable angles between two orthogonal sparse vectors upon compression, this paper shows that the upper bound on** $\delta_{K+1}$ **can be further relaxed to**

$$\delta_{K+1} < \frac{\sqrt{4K+1}-1}{2K}.$$

**This result thus narrows the gap between the so far best known bound and the ultimate performance guarantee** $\delta_{K+1} = \frac{1}{\sqrt{K}}$ **that is conjectured by Dai and Milenkovic in 2009. The proposed approximate orthogonality condition is also exploited to derive less restricted sufficient conditions for signal reconstruction in several compressive sensing problems, including signal recovery via OMP in a noisy environment, compressive domain interference cancellation, and support identification via the subspace pursuit algorithm.**

*Suggested Editorial Area:* **Signal Processing.**

# I. INTRODUCTION

*A. Overview*

   Orthogonal matching pursuit (OMP) [1-4] is a popular greedy approach capable of recovering a *K*-sparse signal $\mathbf{x} \in \mathbb{R}^N$ based on a set of incomplete measurements $\mathbf{y} \in \mathbb{R}^M$ obeying the linear model

$$\mathbf{y} = \mathbf{\Phi}\mathbf{x}, \tag{1.1}$$

This work is sponsored by the National Science Council of Taiwan under grants NSC 102-2221-E-009-019-MY3, NSC 100-2221-E-009-104-MY3 and NSC 100-2628-E-009-008, by the Ministry of Education of Taiwan under the MoE ATU Program, and by the Telecommunication Laboratories, Chunghwa Telecom Co., Ltd. under grant TL-99-G107.

where $\mathbf{\Phi} \in \mathbb{R}^{M \times N}$ is the sensing matrix with $N \gg M (\gg K)$. Basically, OMP (outlined in Table I) is an iterative algorithm, whereby in each iteration an index of the signal support is identified. Under the noiseless system model (1.1), the study of sufficient conditions for perfect signal recovery using OMP recently received considerable attention in the area of compressive sensing [2-4]. Various reports on the performance guarantee of OMP have been available, all of which are specified in terms of either the restricted isometry property (RIP) or the mutual coherence of the sensing matrix $\mathbf{\Phi}$ [4-11]. RIP-based analysis of an OMP-like algorithm, in which the square-error metric for support identification in each iteration is replaced by a general convex objective function, is considered in [12]. Reconstruction of a class of structured sparse signals modeled by trigonometric polynomials via OMP is addressed in [13-14]. A comprehensive review of greedy algorithms for sparse signal recovery, as well as the related analytical performance guarantees, can be found in [3, Chap. 8].

*B. RIP-Based Performance Guarantees: Existing Results*

The sensing matrix $\mathbf{\Phi}$ is said to satisfy RIP of order $K$ [15-17] if there exists $0 < \delta_K < 1$ such that

$$(1 - \delta_K)\|\mathbf{s}\|_2^2 \leq \|\mathbf{\Phi s}\|_2^2 \leq (1 + \delta_K)\|\mathbf{s}\|_2^2 \qquad (1.2)$$

holds for all $K$-sparse $\mathbf{s}$. The constant $\delta_K$ is the so-called restricted isometry constant (RIC) of the sensing matrix $\mathbf{\Phi}$. Under the noiseless model (1.1), Davenport and Wakin [5] showed that the OMP algorithm can exactly identify the support of a $K$-sparse signal in $K$ iterations if $\mathbf{\Phi}$ satisfies RIP of order $K+1$ with RIC $\delta_{K+1} < 1/(3\sqrt{K})$. Hung and Zhu [6] then derived the less restricted sufficient condition $\delta_{K+1} < 1/(1 + \sqrt{2K})$. As the latest report, Mo and Shen [7], and Wang and Shim [8], both proved that the upper bound on $\delta_{K+1}$ can be further relaxed to

$$\delta_{K+1} < \frac{1}{\sqrt{K}+1}. \qquad (1.3)$$

In [7] and [8], examples are also given to illustrate the failure of exact support identification in $K$ iterations in case that

$$\delta_{K+1} = \frac{1}{\sqrt{K}}. \qquad (1.4)$$

Such results also verified the conjecture made by Dai and Milenkovic in [9], viz., values of $\delta_{K+1}$ no less than $1/\sqrt{K}$ may result in the failure of perfect support recovery.

Table I. OMP Algorithm. In Step 3.2, $\mathbf{e}_i$ denotes the *i*th unit standard vector of a suitable dimension; in Step 3.4, the entries of $\mathbf{q}_{\Omega^j} \in \mathbb{R}^j$ are those of the updated $\mathbf{q}$ corresponding to the index set $\Omega^j$.

| |
|---|
| 1. Input: $\mathbf{y}$, $\mathbf{\Phi}$ |
| 2. Initialize: $j = 0$, $\mathbf{r}^0 := \mathbf{y} = \mathbf{\Phi}\mathbf{x}$, $\Omega^0 := [\,]$ and $\mathbf{\Phi}_{\Omega^0} = [\,]$ |
| 3. While $j < K$ <br>     3.1 $j = j+1$ <br>     3.2 $\rho^j = \underset{i=1,\ldots,N}{\arg\max}\left\|\left\langle \mathbf{\Phi}\mathbf{e}_i, \mathbf{r}^{j-1} \right\rangle\right\|$ and $\Omega^j = \Omega^{j-1} \cup \rho^j$ <br>     3.3 $\mathbf{q} = [0\ 0\ \cdots\ 0]^* \in \mathbb{R}^N$ <br>     3.4 $\mathbf{q}_{\Omega^j} = \underset{\mathbf{b}}{\arg\min}\left\|\mathbf{\Phi}_{\Omega^j}\mathbf{b} - \mathbf{y}\right\|_2 = (\mathbf{\Phi}_{\Omega^j}^*\mathbf{\Phi}_{\Omega^j})^{-1}\mathbf{\Phi}_{\Omega^j}^*\mathbf{y}$ <br>     3.5 $\mathbf{r}^j = \mathbf{y} - \mathbf{\Phi}_{\Omega^j}\mathbf{q}_{\Omega^j}$ <br>end while |
| 4. Output: $\hat{\mathbf{x}} = \mathbf{q}$ |

## C. Paper Contribution

In this paper, we show that the upper bound on $\delta_{K+1}$ in the sufficient signal reconstruction condition can be improved even further to

$$\delta_{K+1} < \frac{\sqrt{4K+1} - 1}{2K}. \tag{1.5}$$

Since $\frac{1}{\sqrt{K}+1} < \frac{\sqrt{4K+1}-1}{2K} < \frac{1}{\sqrt{K}}$, our solution thus narrows the gap between the so far best known bound (1.3) and the conjectured ultimate performance guarantee (1.4). Our proof exploits a newly established "approximate orthogonality" condition, which is characterized via achievable angles between two orthogonal sparse vectors upon compression. Hence, as compared to (1.3), the improved performance guarantee (1.5) benefits from more explicit geometric insights of compressed sparse vectors under the RIP of the sensing matrix. The proposed approximate orthogonality condition as well as the proof techniques used for deriving the new bound (1.5) has a far-reaching impact: it finds applications in RIP-based performance analyses for several signal reconstruction problems in compressive sensing. Specifically,

- *Signal Reconstruction via OMP Under Measurement Noise:* In the noisy case, sufficient conditions for exact support identification via OMP are commonly characterized in terms of RIC as well as certain thresholds for the minimal amplitude of the signal components (see, e.g., [18-19]). The proposed approach in our paper can be used for deriving a less restricted sufficient condition as compared to the most recent report in [18]; in particular, it is shown that a less conservative requirement on the RIC of the sensing matrix together with a smaller threshold for the minimal signal magnitude suffices to guarantee exact support recovery.

- *Compressive-Domain Interference Cancellation via Orthogonal Projection [20-24]:* In this problem, a central issue regarding the study of the signal reconstruction performance upon interference removal is to specify the RIC of the effective sensing matrix, which is a product of an orthogonal projection matrix and a random sensing matrix [20-24]. Based on the developed analysis techniques in this paper, we derive a more accurate estimate of the RIC of the effective sensing matrix as compared to the previous works [20] and [24].

- *Support Identification via Subspace Pursuit (SP) [9]:* SP is another popular greedy algorithm for sparse signal recovery in compressive sensing [9], and RIP-based performance guarantee for SP has been investigated in [9] and [10]. By using the proposed approach (in particular, the approximate orthogonality condition), we show in this paper that, to guarantee perfect/stable signal reconstruction via SP, the requirement on the RIC of the sensing matrix $\mathbf{\Phi}$ can be relaxed even further. Specifically, assuming that the sensing matrix $\mathbf{\Phi}$ satisfies RIP of order $3K$, it is shown that $\delta_{3K} \leq 0.2412$ suffices to guarantee exact (stable, respectively) support identification via SP in the noiseless (noisy, respectively) case. Our bound thus improves the results in [9] and [10]: in the absence of noise, the requirement on RIC reported in [9] is $\delta_{3K} < 0.165$; when noise is present, the required bound shown in [9] is $\delta_{3K} < 0.083$, and in [10] is then pushed to $\delta_{3K} < 0.139$.

The organization of this paper is as follows. Section II considers OMP in the noiseless environment, and derives the improved RIC bound (1.5). Section III then focuses on the noisy case, and, under certain noise models, derives less restricted sufficient conditions for OMP to perfectly identify the support. Section IV further investigates the applications of the proposed proof techniques in the study of two other signal reconstruction problems, namely, compressive-domain interference cancellation and signal recovery via SP. For the former, a more accurate estimate of the RIC of the effective sensing matrix upon interference removal is derived; for the later, less restricted sufficient conditions for signal reconstruction are developed. Some concluding remarks are then drawn in Section V. To ease reading, most of the technical proof is relegated to the appendix.

*Notation List:* For $S \subset \{1,\cdots,N\}$ with cardinality $|S|$, we will use $\mathbf{\Phi}_S \in \mathbb{R}^{M \times |S|}$ to denote the matrix obtained from $\mathbf{\Phi} \in \mathbb{R}^{M \times N}$ by retaining its columns indexed by the subset $S$. For $\mathbf{u} \in \mathbb{R}^N$ with $(\mathbf{u})_i$ as the $i$th entry, $\mathbf{u}_S \in \mathbb{R}^{|S|}$ denotes the vector whose entries consist of those of $\mathbf{u}$ indexed by $S$; we write $\tilde{\mathbf{u}}_S \in \mathbb{R}^N$ to be the zero-padded version of $\mathbf{u}_S$ such that $(\tilde{\mathbf{u}}_S)_i = (\mathbf{u})_i$ for $i \in S$ and $(\tilde{\mathbf{u}}_S)_i = 0$ otherwise (thus, $\tilde{\mathbf{u}}_S$ is $|S|$-sparse with support $S$). Throughout the paper, $\mathbf{e}_i \in \mathbb{R}^N$ denotes the $i$th unit standard vector, $\mathbf{I}$ denotes the identity matrix of a proper dimension, $\mathbf{0}$ represents the zero vector of a proper dimension, and $()^*$ stands for the transpose operation. $\mathcal{R}(\mathbf{M})$ denotes the column space of the matrix $\mathbf{M}$. $\|\cdot\|_2$ and $\|\cdot\|_\infty$ denote, respectively, the vector two-norm and sup norm [25]; $\langle \mathbf{u}, \mathbf{v} \rangle$ represents the standard inner product between the two vectors $\mathbf{u}$ and $\mathbf{v}$.

# II. IMPROVED PERFORMANCE GUARANTEE FOR OMP: NOISELESS CASE

*A. Main Result*

In the sequel we denote the support of the desired $K$-sparse vector $\mathbf{x}$ by $T$, with cardinality $|T| = K$. In addition, as in various previous works, e.g., [8], [10-11], it is assumed that all columns of the sensing matrix $\mathbf{\Phi}$ are normalized to be of unit-norm. The development of the main result relies crucially on the next lemma, which characterizes the achievable angle between two compressed orthogonal sparse vectors in terms of the RIC of the sensing matrix.

*Lemma 2.1*: Let $\mathbf{u}$ and $\mathbf{v}$ be two orthogonal sparse vectors with supports $T_u$ and $T_v$ fulfilling $|T_u \cup T_v| \leq K$. Suppose that the sensing matrix $\mathbf{\Phi}$ satisfies RIP of order $K$ with RIC $\delta_K$. Then we have

$$\left|\cos \angle (\mathbf{\Phi}\mathbf{u}, \mathbf{\Phi}\mathbf{v})\right| \leq \delta_K, \tag{2.1}$$

where $\angle(\mathbf{\Phi}\mathbf{u}, \mathbf{\Phi}\mathbf{v})$ denotes the angle between $\mathbf{\Phi}\mathbf{u}$ and $\mathbf{\Phi}\mathbf{v}$.

*[Proof]*: See Appendix A. □

With the aid of Lemma 2.1, we have the following theorem.

*Theorem 2.2*: Assume that the sensing matrix $\mathbf{\Phi}$ satisfies RIP of order $K+1$ with RIC

$$\delta_{K+1} < \frac{\sqrt{4K+1}-1}{2K}. \tag{2.2}$$

Then, for any $K$-sparse $\mathbf{x}$, OMP can perfectly identify the support of $\mathbf{x}$ from the measurement $\mathbf{y} = \mathbf{\Phi}\mathbf{x}$ in $K$ iterations. □

Since $\frac{1}{\sqrt{K+1}} < \frac{\sqrt{4K+1}-1}{2K} < \frac{1}{\sqrt{K}}$, the proposed bound (2.2) thus improves the best-known result $\frac{1}{\sqrt{K+1}}$. Figure 1 shows the three bounds for various sparsity levels $K$. As can be seen from the figure, the improvement over $\frac{1}{\sqrt{K+1}}$ is slight when $K$ is large. This is not unexpected since, for large $K$, the gap between $\frac{1}{\sqrt{K+1}}$ and the conjectured limit $\frac{1}{\sqrt{K}}$ is pretty small, and it is therefore rather difficult to achieve a substantial improvement.

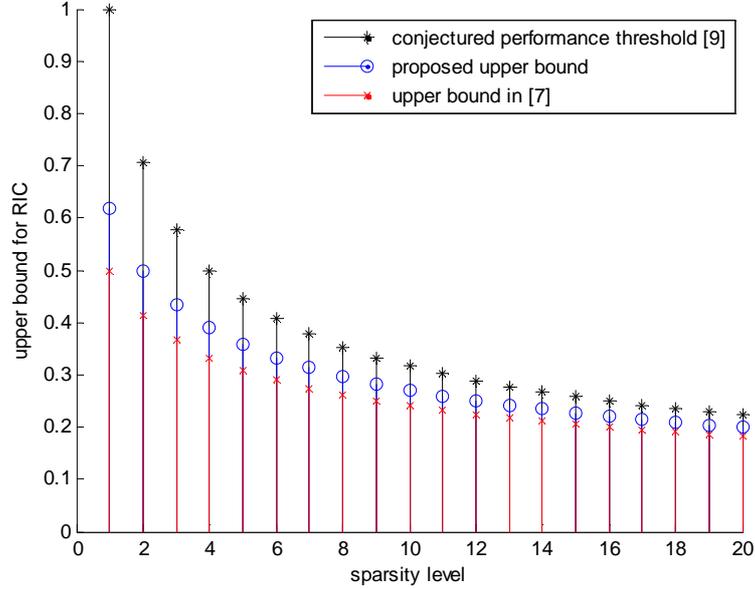

Fig. 1. Comparison of three bounds on $\delta_{K+1}$ for different sparsity levels $K$.

***Remark:*** The derivation of the improved bound (2.2) is built largely on Lemma 2.1, which asserts (in terms of the achievable angle) that the compressed vectors $\mathbf{\Phi u}$ and $\mathbf{\Phi v}$ are nearly orthogonal as long as $\mathbf{u}$ and $\mathbf{v}$ are orthogonal. Notably, under the same assumptions made as in Lemma 2.1 and based on a plane-geometry analysis, the following upper bound on $|\cos\angle(\mathbf{\Phi u},\mathbf{\Phi v})|$ has been derived in [24, p-2071]:

$$|\cos\angle(\mathbf{\Phi u},\mathbf{\Phi v})| \leq \frac{\delta_K}{\sqrt{1-\delta_K^2}}. \qquad (2.3)$$

It can be seen that the proposed bound (2.1), which exploits a geometric interpretation of the two-norm condition number (details referred to Appendix A), improves the result (2.3). An alternative characterization of the "near orthogonality" is via the inner product between $\mathbf{\Phi u}$ and $\mathbf{\Phi v}$ [17]; more precisely, for $\mathbf{u}$ and $\mathbf{v}$ with non-overlapping supports (thus, $\mathbf{u}$ and $\mathbf{v}$ are orthogonal), it has been shown in [17] that

$$|\langle\mathbf{\Phi u},\mathbf{\Phi v}\rangle| \leq \delta_K \|\mathbf{u}\|_2 \cdot \|\mathbf{v}\|_2. \qquad (2.4)$$

It is worthy of noting that (2.4) in conjunction with the RIP condition (1.2) can be directly used to derive an upper bound of $|\cos\angle(\mathbf{\Phi u},\mathbf{\Phi v})|$, as can be seen by

$$|\cos\angle(\mathbf{\Phi u},\mathbf{\Phi v})| = \frac{|\langle\mathbf{\Phi u},\mathbf{\Phi v}\rangle|}{\|\mathbf{\Phi u}\|_2 \cdot \|\mathbf{\Phi v}\|_2} \stackrel{(a)}{\leq} \frac{\delta_K \|\mathbf{u}\|_2 \cdot \|\mathbf{v}\|_2}{\|\mathbf{\Phi u}\|_2 \cdot \|\mathbf{\Phi v}\|_2} \stackrel{(b)}{\leq} \frac{\delta_K \|\mathbf{u}\|_2 \cdot \|\mathbf{v}\|_2}{(1-\delta_K)\|\mathbf{u}\|_2 \cdot \|\mathbf{v}\|_2} = \frac{\delta_K}{1-\delta_K}, \qquad (2.5)$$

where (a) follows from (2.4) and (b) holds due to the RIP (1.2). The upper bound (2.5) derived by using the simple algebraic approach shown above is even worse than (2.3); this is not unexpected since algebraic analyses using RIP are known to yield the worst-case estimate [26-27]. To summarize, Lemma 2.1 asserts that, when $\mathbf{u}$ and $\mathbf{v}$ are orthogonal, the measure of orthogonality between $\mathbf{\Phi u}$ and $\mathbf{\Phi v}$ in terms of the achievable $\angle(\mathbf{\Phi u}, \mathbf{\Phi v})$ can be improved as compared to the previous results given by (2.3) and (2.5). We would like to mention that the "near orthogonality" condition (in the form of (2.1) or (2.4)) plays a fundamental role in the study of the signal reconstruction performance in compressive sensing [17], [21], [24]. Thanks to the improved bound (2.1), less restricted sufficient conditions for several signal reconstruction schemes can then be obtained; the related study is given in Sections III and IV.

*B. Detailed Proof of Theorem 2.2*

Note that, in the $j$-th iteration, the index $\rho^j$ determined yields the maximal $\left|\left\langle \mathbf{\Phi e}_i, \mathbf{r}^{j-1} \right\rangle\right|$ (see Step 3.2 of the algorithm). We claim that, under the assumption (2.2), such $\rho^j$, where $1 \leq j \leq K$, belongs to the support $T$. Also, according to the orthogonality property of OMP (see, e.g., [8, Lemma 7]), all the already selected atoms will not be selected again and, thus, all the selected indexes $\rho^j$'s, $j = 1, \cdots, K$, are distinct. The assertion of the theorem then follows.

To prove the claim, it suffices to show that, for each $1 \leq j \leq K$, the following two conditions hold:

$$\left|\left\langle \mathbf{\Phi e}_i, \mathbf{r}^{j-1} \right\rangle\right| \leq \left\|\mathbf{r}^{j-1}\right\|_2 \delta_{K+1} \quad \text{for all } i \notin T, \tag{2.6}$$

and

$$\left|\left\langle \mathbf{\Phi e}_i, \mathbf{r}^{j-1} \right\rangle\right| > \left\|\mathbf{r}^{j-1}\right\|_2 \delta_{K+1} \quad \text{for some } i \in T, \tag{2.7}$$

where $\mathbf{\Phi e}_i \in \mathbb{R}^M$ is the $i$th column of $\mathbf{\Phi}$. Toward this end, we first rewrite $\left|\left\langle \mathbf{\Phi e}_i, \mathbf{r}^{j-1} \right\rangle\right|$ as

$$\left|\left\langle \mathbf{\Phi e}_i, \mathbf{r}^{j-1} \right\rangle\right| = \left\|\mathbf{\Phi e}_i\right\|_2 \left\|\mathbf{r}^{j-1}\right\|_2 \left|\cos \angle \left(\mathbf{\Phi e}_i, \mathbf{r}^{j-1}\right)\right| \stackrel{(a)}{=} \left\|\mathbf{r}^{j-1}\right\|_2 \left|\cos \angle \left(\mathbf{\Phi e}_i, \mathbf{r}^{j-1}\right)\right|, \tag{2.8}$$

where (a) holds since by assumption each column of $\mathbf{\Phi}$ is with unit norm. Based on (2.8), the proof will be done by induction.

In the first iteration ($j = 1$), $\mathbf{r}^{j-1}$ needed for computing the inner product in Step 3.2 of the OMP algorithm is

$$\mathbf{r}^0 = \mathbf{y} = \mathbf{\Phi x}, \tag{2.9}$$

where $\mathbf{x}$ is the desired $K$-sparse vector supported on $T$. Hence, as long as $i \notin T$, we have $\langle \mathbf{e}_i, \mathbf{x} \rangle = 0$ and $|\{i\} \cup T| = K + 1$. According to Lemma 2.1, it follows immediately that

$$\left|\cos \angle \left(\mathbf{\Phi e}_i, \mathbf{r}^0\right)\right| = \left|\cos \angle \left(\mathbf{\Phi e}_i, \mathbf{\Phi x}\right)\right| \leq \delta_{K+1} \quad \text{for all } i \notin T. \tag{2.10}$$

The assertion of (2.6) for $j = 1$ thus follows from (2.8) and (2.10). To prove that (2.7) is true for $j = 1$,

assume otherwise that

$$\left|\langle \mathbf{\Phi e}_i, \mathbf{r}^0 \rangle\right| \leq \left\|\mathbf{r}^0\right\|_2 \delta_{K+1} \ \ \text{for all} \ \ i \in T. \tag{2.11}$$

We will then show that (2.11) leads to contradiction. Let us write $\mathbf{\Phi x} = \sum_{i \in T} (\mathbf{x})_i \mathbf{\Phi e}_i$. Accordingly, we have

$$\begin{aligned}
\|\mathbf{\Phi x}\|_2 &= \frac{|\langle \mathbf{\Phi x}, \mathbf{\Phi x} \rangle|}{\|\mathbf{\Phi x}\|_2} \stackrel{(a)}{=} \frac{|\langle \mathbf{\Phi x}, \mathbf{r}^0 \rangle|}{\|\mathbf{r}^0\|_2} = \frac{\left|\langle \sum_{i \in T}(\mathbf{x})_i \mathbf{\Phi e}_i, \mathbf{r}^0 \rangle\right|}{\|\mathbf{r}^0\|_2} = \frac{\left|\sum_{i \in T}(\mathbf{x})_i \langle \mathbf{\Phi e}_i, \mathbf{r}^0 \rangle\right|}{\|\mathbf{r}^0\|_2} \\
&\leq \frac{\sum_{i \in T} |(\mathbf{x})_i| \left|\langle \mathbf{\Phi e}_i, \mathbf{r}^0 \rangle\right|}{\|\mathbf{r}^0\|_2} \\
&\stackrel{(b)}{\leq} \frac{\sum_{i \in T} |(\mathbf{x})_i| \|\mathbf{r}^0\|_2 \delta_{K+1}}{\|\mathbf{r}^0\|_2} = \|\mathbf{x}\|_1 \delta_{K+1} \\
&\stackrel{(c)}{\leq} \sqrt{K} \|\mathbf{x}\|_2 \delta_{K+1} \\
&\stackrel{(d)}{<} \sqrt{K} \|\mathbf{x}\|_2 \frac{\sqrt{4K+1}-1}{2K} = \|\mathbf{x}\|_2 \frac{\sqrt{4K+1}-1}{2\sqrt{K}},
\end{aligned} \tag{2.12}$$

where (a) follows from (2.9), (b) follows from the assumption (2.11), (c) holds since $\mathbf{x}$ is $K$-sparse and, as a result, $\|\mathbf{x}\|_1 \leq \sqrt{K}\|\mathbf{x}\|_2$, and (d) is due to (2.2). Again, by using (2.2), we can obtain the following inequality

$$1 - \delta_{K+1} > \frac{2K+1-\sqrt{4K+1}}{2K} \cdot \frac{2K}{2K} = \frac{4K^2 + 2K - 2K\sqrt{4K+1}}{2K \cdot 2K} = \frac{K(\sqrt{4K+1}-1)^2}{(2K)^2}, \tag{2.13}$$

or equivalently,

$$\sqrt{1 - \delta_{K+1}} > \frac{\sqrt{K}(\sqrt{4K+1}-1)}{2K} = \frac{\sqrt{4K+1}-1}{2\sqrt{K}}. \tag{2.14}$$

Combining (2.12) and (2.14) yields

$$\|\mathbf{\Phi x}\|_2 < \sqrt{1 - \delta_{K+1}} \|\mathbf{x}\|_2, \tag{2.15}$$

which contradicts with the RIP condition $\|\mathbf{\Phi x}\|_2 \geq \sqrt{1 - \delta_{K+1}} \|\mathbf{x}\|_2$. Hence, not only (2.6) but also (2.7) holds for $j = 1$.

Let us proceed into the second iteration ($j = 2$). Since (2.6) and (2.7) are true for $j = 1$, we have $\rho^1 \in T$, that is, the index selected in the first iteration belongs to the support $T$; it is noted again that the index $\rho^1$ determined in the first iteration will not be selected again due to the orthogonality property of OMP [8, Lemma 7]. Since $\rho^1 \in T$, the residual vector $\mathbf{r}^1$ admits the sparse representation (see the proof of Lemma 6 in [8, p-4975])

$$\mathbf{r}^1 = \mathbf{\Phi z} \tag{2.16}$$

for some $K$-sparse $\mathbf{z}$ also supported on $T$. With the aid of (2.16) and by following the similar arguments for deriving (2.10), it can be shown that $\left|\cos\angle\left(\mathbf{\Phi}\mathbf{e}_i,\mathbf{r}^1\right)\right|=\left|\cos\angle\left(\mathbf{\Phi}\mathbf{e}_i,\mathbf{\Phi}\mathbf{z}\right)\right|\leq\delta_{K+1}$ for all $i\notin T$: it thus follows from (2.8) that (2.6) is true when $j=2$. Also, based on essentially the same contradiction-based arguments (i.e., starting from (2.11) with $\mathbf{r}^0$ replaced by $\mathbf{r}^1$, and then in (2.12) with $\mathbf{x}$ and $\mathbf{r}^0$ replaced by, respectively, $\mathbf{z}$ and $\mathbf{r}^1$), it can be shown that (2.7) is true for $j=2$. We then proceed into the third iteration ($j=3$). Since (2.6) and (2.7) are true for $j=2$, it follows $\rho^2\in T$. Now, we have $\rho^1,\rho^2\in T$. Again, according to [8, p-4975], the residual vector $\mathbf{r}^2$ admits a $K$-sparse representation of the form (2.16), with some $\mathbf{z}$ supported on $T$. By repeating the same proof procedures it can be shown that (2.6) and (2.7) are true for $j=3$. By continuing this process, it can be concluded that (2.6) and (2.7) hold for $j=4,\cdots,K$. The assertion of the claim is thus proved.

## III. OMP BASED SUPPORT IDENTIFICATION WITH NOISE

To guarantee exact support identification via OMP in the presence of noise[1], sufficient conditions specified in terms of the RIC $\delta_{K+1}$ and certain signal amplitude thresholds have been reported in [18]. The improved upper bound (2.2) on $\delta_{K+1}$ together with Lemma 2.1 can be exploited to derive less restricted sufficient conditions, as will be shown below.

Now, we consider the following signal model under noise corruption

$$\mathbf{y}=\mathbf{\Phi}\mathbf{x}+\mathbf{w}, \qquad (3.1)$$

where $\mathbf{w}\in\mathbb{R}^M$ is the measurement noise. The next proposition states the sufficient conditions developed in [18]. Through this section, $\mathbf{r}^j$ denotes the residual vector in the $j$-th iteration of the OMP algorithm (cf. Step 3.5 in Table I).

***Proposition 3.1 [18]:*** Consider the signal model (3.1). Then the following results hold.

(1). ($\ell_2$-bounded noise) Under $\|\mathbf{w}\|_2\leq\varepsilon_1$, OMP with the stopping criterion $\left\|\mathbf{r}^j\right\|_2\leq\varepsilon_1$ can exactly identify the support $T$ of the $K$-sparse signal $\mathbf{x}$ if $\delta_{K+1}<1/(\sqrt{K}+1)$ and the minimum magnitude of the nonzero entries of $\mathbf{x}$ satisfies

$$\min_{i\in T}\left|(\mathbf{x})_i\right|>\frac{\left(\sqrt{1+\delta_{K+1}}+1\right)\varepsilon_1}{1-\delta_{K+1}-\sqrt{K}\delta_{K+1}}. \qquad (3.2)$$

---

1. When noise is present, some stopping criterion in terms of the norm of the residual vector should be added to terminate the algorithm [18].

(2). ($\ell_\infty$-bounded noise) Under $\|\mathbf{\Phi}^*\mathbf{w}\|_\infty \leq \varepsilon_2$, OMP with the stopping criterion $\|\mathbf{\Phi}^*\mathbf{r}^j\|_\infty \leq \varepsilon_2$ can exactly identify the support $T$ of the $K$-sparse signal $\mathbf{x}$ if $\delta_{K+1} < 1/(\sqrt{K}+1)$ and the minimum magnitude of the nonzero entries of $\mathbf{x}$ satisfies

$$\min_{i \in T} |(\mathbf{x})_i| > \frac{\left(\sqrt{K} + \sqrt{K}\sqrt{1+\delta_{K+1}}\right)\varepsilon_2}{1-\delta_{K+1} - \sqrt{K}\delta_{K+1}}. \tag{3.3}$$

□

Thanks to the improved upper bound in (2.2) and the improved approximate orthogonality condition (2.1), less restricted sufficient conditions for exact support identification via OMP under noise corruption are shown in the next theorem.

***Theorem 3.2:*** Consider the signal model (3.1). Then the following results hold.

(1). ($\ell_2$-bounded noise) Under $\|\mathbf{w}\|_2 \leq \varepsilon_1$, OMP with the stopping criterion $\|\mathbf{r}^j\|_2 \leq \varepsilon_1$ can exactly identify the support $T$ of the $K$-sparse signal $\mathbf{x}$ if $\delta_{K+1}$ fulfills (2.2) and the minimum magnitude of the nonzero entries of $\mathbf{x}$ satisfies

$$\min_{i \in T} |(\mathbf{x})_i| > \frac{(\sqrt{1+\delta_{K+1}}+1)\varepsilon_1}{1-\delta_{K+1} - \sqrt{1-\delta_{K+1}}\sqrt{K}\delta_{K+1}}. \tag{3.4}$$

(2). ($\ell_\infty$-bounded noise) Under $\|\mathbf{\Phi}^*\mathbf{w}\|_\infty \leq \varepsilon_2$, OMP with the stopping criterion $\|\mathbf{\Phi}^*\mathbf{r}^j\|_\infty \leq \varepsilon_2$ can exactly identify the support $T$ of the $K$-sparse signal $\mathbf{x}$ if $\delta_{K+1}$ fulfills (2.2) and the minimum magnitude of the nonzero elements of $\mathbf{x}$ satisfies

$$\min_{i \in T} |(\mathbf{x})_i| > \frac{(\sqrt{K}+1)\varepsilon_2}{1-\delta_{K+1} - \sqrt{1-\delta_{K+1}}\sqrt{K}\delta_{K+1}}. \tag{3.5}$$

*[Proof]:* The proof basically follows the induction argument as in the proof of Theorem 2.2, and the details are referred to Appendix B. □

For either noise model, it can be readily checked that the derived threshold (i.e., (3.4) or (3.5)) is smaller than that shown in [18] (i.e., (3.2) or (3.3)). Our results thus assert that, to guarantee exact support identification via OMP in a noisy case, the requirements on the RIC and the minimal signal amplitude can be further relaxed as compared to the report in [18].

***Remark:*** For the Gaussian noise case, $\|\mathbf{w}\|_2$ can be bounded from above with a sufficiently high probability. Hence, under an additional probability constraint, an improved minimal magnitude threshold for the Gaussian case can also be obtained by following essentially the same procedures as in the $\ell_2$-bounded noise case. □

# IV. IMPACTS OF THE APPROXIMATE ORTHOGONALITY CONDITION (2.1)

As mentioned above, the approximate orthogonality condition (2.1) (measured in terms of achievable angles) not only enjoys its own technical novelty but also has a wide spectrum of applications: it can be used for developing improved RIP-based performance characterizations for other sparse signal reconstruction schemes. Below we discuss two such applications in details.

*A. Sparse Signal Recovery Against Sparse Interference via Orthogonal Projection*

Consider the following compressive sensing system under sparse interference corruption [20]

$$\mathbf{y} = \mathbf{\Phi}(\mathbf{x} + \mathbf{d}) = \mathbf{\Phi}\mathbf{x} + \mathbf{\Phi}\mathbf{d}, \qquad (4.1)$$

where $\mathbf{d} \in \mathbb{R}^N$ is a sparse interference/disturbance with support $T_d$. As in various previous works [20-24], it is assumed that $T_d$ is known and does not overlap with the signal support $T$. To exploit the knowledge of $T_d$ for interference removal, one typical approach is via orthogonal projection. More specifically, the measurement $\mathbf{y}$ is projected onto the orthogonal complement of the interference subspace $\mathcal{R}(\mathbf{\Phi}_{T_d})$ to obtain [20-24]

$$\overline{\mathbf{y}} \triangleq \mathbf{Q}\mathbf{y} = \mathbf{Q}\mathbf{\Phi}\mathbf{x} + \mathbf{Q}\mathbf{\Phi}\mathbf{d} = \mathbf{Q}\mathbf{\Phi}\mathbf{x}, \qquad (4.2)$$

where the projection matrix $\mathbf{Q} \triangleq \mathbf{I} - \mathbf{\Phi}_{T_d}(\mathbf{\Phi}^*_{T_d}\mathbf{\Phi}_{T_d})^{-1}\mathbf{\Phi}^*_{T_d}$ removes all the components of $\mathbf{y}$ lying in $\mathcal{R}(\mathbf{\Phi}_{T_d})$. Upon interference removal, the effective sensing matrix in (4.2) is $\mathbf{Q}\mathbf{\Phi}$, which is a product of an orthogonal projection matrix $\mathbf{Q}$ and the original random sensing matrix $\mathbf{\Phi}$. The performance of sparse signal reconstruction based on (4.2) depends crucially on the RIC of $\mathbf{Q}\mathbf{\Phi}$. The RIP of $\mathbf{Q}\mathbf{\Phi}$ as well as the achievable RIC was first studied in [20]; the results are summarized in the next proposition.

***Proposition 4.1 [20]:*** Consider the system (4.2). Assume that $\mathbf{\Phi}$ satisfies the RIP of order $K$ with RIC given by $\delta_K$, and that the interference support $T_d$ satisfies $|T_d| < K$. The following inequality holds for all $(K - |T_d|)$-sparse $\mathbf{x}$ whose support does not overlap with $T_d$:

$$\left(1 - \overline{\delta}_A\right)\|\mathbf{x}\|_2^2 \leq \|\mathbf{Q}\mathbf{\Phi}\mathbf{x}\|_2^2 \leq (1 + \delta_K)\|\mathbf{x}\|_2^2, \qquad (4.3)$$

where

$$\overline{\delta}_A \triangleq \min\left\{1, \frac{\delta_K}{1 - \delta_K}\right\}. \qquad (4.4)$$

□

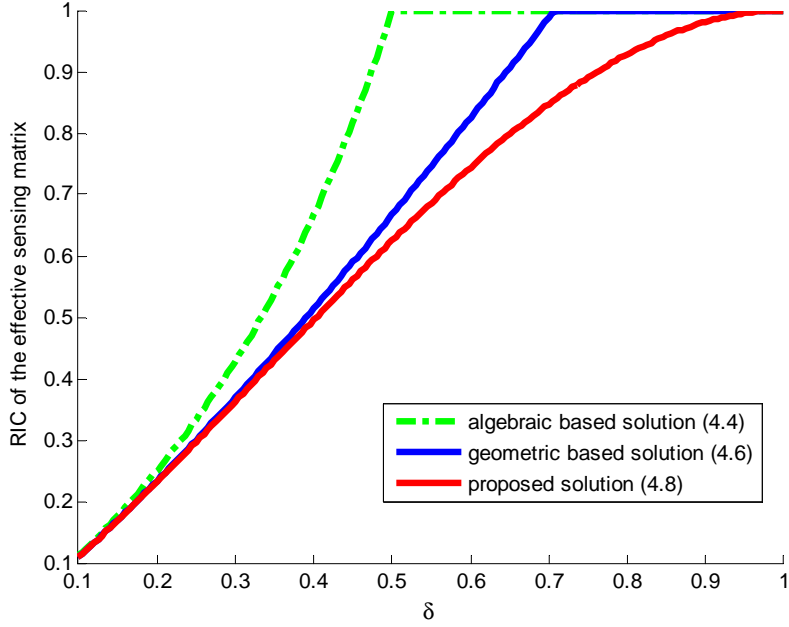

Fig. 2 RIC of $\mathbf{P\Phi}$ for different values of $\delta$, the RIC of $\mathbf{\Phi}$.

Proposition 4.1 asserts that $\mathbf{Q\Phi}$ still enjoys RIP, but is with an RIC $\overline{\delta}_A$ larger than $\delta_K$. In [24], an improved estimate of RIC of $\mathbf{Q\Phi}$ was obtained by means of certain plane geometry analyses, as asserted in the next proposition.

***Proposition 4.2 [24]:*** Under the same assumptions as in Proposition 4.1, the following inequality holds for all $(K-|T_d|)$-sparse $\mathbf{x}$ whose support does not overlap with $T_d$:

$$\left(1-\overline{\delta}_G\right)\|\mathbf{x}\|_2^2 \leq \|\mathbf{Q\Phi x}\|_2^2 \leq (1+\delta_K)\|\mathbf{x}\|_2^2, \tag{4.5}$$

where

$$\overline{\delta}_G \triangleq \min\left\{1, \delta_K + \frac{\delta_K^2}{1+\delta_K}\right\}. \tag{4.6}$$

By leveraging Lemma 2.1, the following theorem shows that the estimate of the RIC of $\mathbf{Q\Phi}$ can be improved even further.

***Theorem 4.3:*** Under the same assumptions as in Proposition 4.1, the following inequality holds for all $(K-|T_d|)$-sparse $\mathbf{x}$ whose support does not overlap with $T_d$:

$$\left(1-\overline{\delta}\right)\|\mathbf{x}\|_2^2 \leq \|\mathbf{Q\Phi x}\|_2^2 \leq (1+\delta_K)\|\mathbf{x}\|_2^2, \tag{4.7}$$

where

$$\overline{\delta} \triangleq \min\left\{1, \delta_K + \delta_K^2(1-\delta_K)\right\}. \tag{4.8}$$

*[Proof]:* See Appendix C.

Table II. The SP Algorithm.

| |
|---|
| 1. Input: $\mathbf{y}$, $\mathbf{\Phi}$ |
| 2. Initialize: <br>     2.1   $\mathbf{r}^0 = \mathbf{y}$ <br>     2.2   $\mathbf{\Omega}^1 = \{K \text{ indexes yielding the largest magnitude entries of } \mathbf{\Phi}^* \mathbf{r}^0 \}$, <br>     2.3   $\mathbf{q} = [0 \; 0 \; \cdots \; 0]^* \in \mathbb{R}^N$ <br>     2.4   $\mathbf{q}_{\mathbf{\Omega}^1} = \arg\min_{\mathbf{b}} \|\mathbf{\Phi}_{\mathbf{\Omega}^1} \mathbf{b} - \mathbf{y}\|_2 = (\mathbf{\Phi}^*_{\mathbf{\Omega}^1} \mathbf{\Phi}_{\mathbf{\Omega}^1})^{-1} \mathbf{\Phi}^*_{\mathbf{\Omega}^1} \mathbf{y}$ <br>     2.5   $\mathbf{r}^1 := \mathbf{y} - \mathbf{\Phi}_{\mathbf{\Omega}^1} \mathbf{q}_{\mathbf{\Omega}^1}$ <br>     2.6   $j = 1$ |
| 3. While $\|\mathbf{r}^{j-1}\|_2 < \|\mathbf{r}^j\|_2$ <br>     3.0   $j = j+1$ <br>     3.1   $\mathbf{\Omega}^j_\Delta = \{K \text{ indexes yielding the largest magnitude entries of } \mathbf{\Phi}^* \mathbf{r}^{j-1}\}$ <br>     3.2   $\breve{\mathbf{\Omega}}^j \triangleq \mathbf{\Omega}^j_\Delta \cup \mathbf{\Omega}^{j-1}$ <br>     3.3   $\breve{\mathbf{q}} = [0 \; 0 \; \cdots \; 0]^* \in \mathbb{R}^N$ <br>     3.4   $\breve{\mathbf{q}}_{\breve{\mathbf{\Omega}}^j} = \arg\min_{\mathbf{b}} \|\mathbf{\Phi}_{\breve{\mathbf{\Omega}}^j} \mathbf{b} - \mathbf{y}\|_2 = (\mathbf{\Phi}^*_{\breve{\mathbf{\Omega}}^j} \mathbf{\Phi}_{\breve{\mathbf{\Omega}}^j})^{-1} \mathbf{\Phi}^*_{\breve{\mathbf{\Omega}}^j} \mathbf{y}$ <br>     3.5   $\mathbf{\Omega}^j = \{K \text{ indexes yielding the largest magnitude entries of } \breve{\mathbf{q}}\}$ <br>     3.6   $\mathbf{q} = [0 \; 0 \; \cdots \; 0]^* \in \mathbb{R}^N$ <br>     3.7   $\mathbf{q}_{\mathbf{\Omega}^j} = \arg\min_{\mathbf{b}} \|\mathbf{\Phi}_{\mathbf{\Omega}^j} \mathbf{b} - \mathbf{y}\|_2 = (\mathbf{\Phi}^*_{\mathbf{\Omega}^j} \mathbf{\Phi}_{\mathbf{\Omega}^j})^{-1} \mathbf{\Phi}^*_{\mathbf{\Omega}^j} \mathbf{y}$ <br>     3.8   $\mathbf{r}^j \triangleq \mathbf{y} - \mathbf{\Phi}_{\mathbf{\Omega}^j} \mathbf{q}_{\mathbf{\Omega}^j} = \mathbf{y} - \mathbf{\Phi}\mathbf{q}$ <br> end while |
| 4. Output: $\hat{\mathbf{x}} = \mathbf{q}$ |

It is easy to verify that $\bar{\delta} < \bar{\delta}_G < \bar{\delta}_A$, viz., the proposed solution (4.8) improves the previous estimates in [20] and [24] (this is also confirmed by Figure 2, which plots the three estimated RIC with respect to different values of $\delta_K$, the RIC of the random sensing matrix $\mathbf{\Phi}$). Since a smaller RIC results in a better signal reconstruction performance [3], our result implies that the achievable performance of sparse signal recovery with interference-nulling is actually better than as predicted by the previous works [20] and [24].

*B. Support Identification via SP*

SP is another popular greedy algorithm for sparse signal recovery in the area of compressive sensing [9]; see Table II for an outline of the algorithm. In each iteration, SP tries to keep track of an estimated support consisting of $K$ elements by adding and then removing certain elements to and from the candidate set. RIP-based performance guarantees for SP, in both noiseless and noisy cases, have been reported in [9], [10]. The following proposition summarizes the result in [9] when noise is absent. In the sequel, $\mathbf{r}^j$ denotes the residual vector in the $j$-th iteration of the SP algorithm (cf. Step 3.8 in Table II).

***Proposition 4.4 [9]:*** Assume that the sensing matrix $\Phi$ satisfies RIP of order $3K$ with RIC

$$\delta_{3K} < 0.165. \tag{4.9}$$

Then, for any $K$-sparse $\mathbf{x}$, the SP algorithm with stopping criterion $\left\|\mathbf{r}^j\right\|_2 \geq \left\|\mathbf{r}^{j-1}\right\|_2$ can perfectly identify the support of $\mathbf{x}$ from the measurement $\mathbf{y} = \Phi\mathbf{x}$ via a finite number of iterations. □

By means of Lemma 2.1, an improved result (in terms of a less strict requirement on the RIC of the sensing matrix) is derived in the next theorem.

***Theorem 4.5:*** Assume that the sensing matrix $\Phi$ satisfies RIP of order $3K$ with RIC

$$\delta_{3K} \leq 0.2412. \tag{4.10}$$

Then, for any $K$-sparse $\mathbf{x}$, the SP algorithm with stopping criterion $\left\|\mathbf{r}^j\right\|_2 \geq \left\|\mathbf{r}^{j-1}\right\|_2$ can perfectly identify the support of $\mathbf{x}$ from the measurement $\mathbf{y} = \Phi\mathbf{x}$ via a finite number of iterations.
*[Proof]:* See Appendix D. □

When noise is present, SP is capable of achieving stable signal reconstruction, in the sense that, if the sensing matrix satisfies RIP with a small RIC, the reconstruction error is bounded and the size does not exceed a constant times the noise level. The following proposition, which is established in [9], makes this point precise.

***Proposition 4.6 [9]:*** Assume that the sensing matrix $\Phi$ satisfies RIP of order $3K$ with RIC

$$\delta_{3K} < 0.083. \tag{4.11}$$

Then, the SP algorithm reconstructs the $K$-sparse vector $\mathbf{x}$ from the measurement $\mathbf{y} = \Phi\mathbf{x} + \mathbf{w}$ with the reconstruction error bounded as

$$\|\mathbf{x} - \hat{\mathbf{x}}\|_2 \leq c'_K \|\mathbf{w}\|_2, \text{ with } c'_K \triangleq \frac{1 + \delta_{3K} + \delta_{3K}^2}{\delta_{3K}(1 - \delta_{3K})}, \tag{4.12}$$

where $\hat{\mathbf{x}}$ is the estimated sparse signal vector. □

By using the variation of the proof of [9, Theorem 10], an improved performance guarantee has been derived in [10], and is given in the next proposition.

***Proposition 4.7 [10]:*** Assume that the sensing matrix $\Phi$ satisfies RIP of order $3K$ with RIC

$$\delta_{3K} < 0.139. \tag{4.13}$$

Then, the SP algorithm reconstructs the $K$-sparse vector $\mathbf{x}$ from the measurement $\mathbf{y} = \mathbf{\Phi}\mathbf{x} + \mathbf{w}$ with the reconstruction error bounded as

$$\|\mathbf{x} - \hat{\mathbf{x}}\|_2 \leq \overline{c}_K \left\|\mathbf{\Phi}^*_{T_e}\mathbf{w}\right\|_2, \text{ with } \overline{c}_K \triangleq 2 \cdot \frac{7 - 9\delta_{3K} + 7\delta_{3K}^2 - \delta_{3K}^3}{(1 - \delta_{3K})^4}, \tag{4.14}$$

where $\hat{\mathbf{x}}$ is the estimated sparse signal vector and $T_e \triangleq \underset{S \text{ with } |S|=K}{\arg\max} \left\|\mathbf{\Phi}^*_S \mathbf{w}\right\|_2$. □

By exploiting the approximate orthogonality property shown in Lemma 2.1, we can obtain a less conservative sufficient condition for guaranteeing stable signal reconstruction as well as a tighter reconstruction error bound. Specifically, we have the following theorem.

***Theorem 4.8:*** Assume that sensing matrix $\mathbf{\Phi}$ satisfies RIP of order $3K$ with RIC

$$\delta_{3K} \leq 0.2412. \tag{4.15}$$

Then, the SP algorithm reconstructs the $K$-sparse vector $\mathbf{x}$ from the measurement $\mathbf{y} = \mathbf{\Phi}\mathbf{x} + \mathbf{w}$ with the reconstruction error bounded as

$$\|\mathbf{x} - \hat{\mathbf{x}}\|_2 \leq c_K \|\mathbf{w}\|_2, \text{ with } c_K \triangleq (1 + \frac{\delta_{3K}\sqrt{1+\delta_{3K}}}{\sqrt{1-\delta_{3K}}}) \frac{(2+\sqrt{1+\delta_{3K}}\beta)}{\sqrt{1-\delta_{3K}} - \sqrt{1+\delta_{3K}}\alpha} + \frac{1}{\sqrt{1-\delta_{3K}}}, \tag{4.16}$$

where $\hat{\mathbf{x}}$ is the estimated sparse signal vector,

$$\alpha \triangleq \frac{2\delta_{3K}}{1-\delta_{3K}} \sqrt{1 + \delta_{3K}^2 \frac{1+\delta_{3K}}{1-\delta_{3K}}} \sqrt{1 + 4\delta_{3K}^2 \frac{1+\delta_{3K}}{1-\delta_{3K}}}, \tag{4.17}$$

and

$$\beta \triangleq \frac{2\sqrt{1+\delta_{3K}}}{1-\delta_{3K}} \sqrt{1 + \frac{4\delta_{3K}^2(1+\delta_{3K})}{1-\delta_{3K}}} + \frac{2}{\sqrt{1-\delta_{3K}}}. \tag{4.18}$$

*[Proof]:* See Appendix E. □

In summary, for either the noiseless or noisy case, we have pushed the bound for $\delta_{3K}$ in the sufficient conditions to $0.2412$. In addition, when noise is present, it can be shown through some algebra that the proposed reconstruction error upper bound (4.16) is smaller than (4.12) (under a fixed RIC $\delta_{3K}$). To further compare our error bound (4.16) with the result in [10], we first use (4.14) to obtain the following bound independent of $\mathbf{\Phi}^*_{T_e}$:

$$\|\mathbf{x} - \hat{\mathbf{x}}\|_2 \leq \overline{c}_K \left\|\mathbf{\Phi}^*_{T_e}\mathbf{w}\right\|_2 \leq \overline{c}_K \sqrt{1+\delta_{3K}} \|\mathbf{w}\|_2, \tag{4.19}$$

where the second inequality can be obtained by using Lemma A.1 in the appendix. Based on (4.19) and by invoking the definition of $\overline{c}_K$ in (4.14), it can be shown that $c_K < \overline{c}_K\sqrt{1+\delta_{3K}}$, viz., the proposed bound (4.16) is also smaller than (4.19). Hence, our analysis shows that the reconstruction performance of SP is actually better than as predicted in [9] and [10].

## V. CONCLUDING REMARKS

In this paper, we first derive an improved RIP-based performance guarantee for perfect support identification via OMP in the noiseless setting. Our result narrows the gap between the so-far best known bound on the RIC of the sensing matrix and the conjectured ultimate threshold. The proposed approach exploits a newly established approximate orthogonality condition, which is characterized via the achievable angles between two compressed orthogonal sparse vectors under RIP. Such a near-orthogonality property in conjunction with the developed analysis techniques evidenced a wider spectrum of applications. Indeed, under the theme behind our analysis, less restricted sufficient conditions for perfect support identification via OMP in a noisy case is further derived. Compared to the most recent work [18], our result shows that a larger RIC together with a smaller threshold on the minimal signal amplitude can ensure exact support identification. Then, we consider the problem of compressive-domain interference cancellation, and derive a more accurate estimate of the RIC of the effective sensing matrix (in comparison to the results in [20] and [24]). Finally, we study support identification via SP in both noiseless and noisy settings. By means of the approximate orthogonality condition, it is shown that, compared to [9] and [10], the requirement on the RIC of the sensing matrix for guaranteeing exact/stable signal recovery can be further relaxed; in addition, when noise is present, a smaller upper bound of the reconstruction error is obtained. Improved RIP-based performance analysis of other greedy algorithms, such as CoSaMP [28], by using the developed analysis techniques in this paper is currently under investigation.

## APPENDIX

*A. Proof of Lemma 2.1*

To prove Lemma 2.1, we need the following two lemmas.

***Lemma A.1 [9]:*** Assume that $\mathbf{\Phi} \in \mathbb{R}^{M \times N}$ satisfies RIP of order $K$ with RIC $\delta_K$. Then, for $S \subset \{1,\ldots,N\}$ with $|S| \leq K$, the $M \times |S|$ sub-matrix $\mathbf{\Phi}_S$ is of full column rank with singular values $\sigma_1(\mathbf{\Phi}_S) \geq \sigma_2(\mathbf{\Phi}_S) \geq \cdots \geq \sigma_{|S|}(\mathbf{\Phi}_S) > 0$ satisfying

$$\sqrt{1-\delta_K} \leq \sigma_j(\mathbf{\Phi}_S) \leq \sqrt{1+\delta_K}, \quad j=1,\cdots,|S|. \tag{A.1}$$

□

***Lemma A.2:*** Let $\mathbf{A} \in \mathbb{R}^{m \times n}$, with $m \geq n$, be of full column rank, and $\sigma_1(\mathbf{A}) \geq \sigma_2(\mathbf{A}) \geq \cdots \geq \sigma_n(\mathbf{A}) > 0$ be the singular values of $\mathbf{A}$. Let $\kappa(\mathbf{A}) = \sigma_1(\mathbf{A})/\sigma_n(\mathbf{A})$ be the condition number of $\mathbf{A}$. Then we have

$$\kappa(\mathbf{A}) = \cot(\theta(\mathbf{A})/2) \tag{A.2}$$

where $0 < \theta(\mathbf{A}) \leq \pi/2$ is given by

$$\theta(\mathbf{A}) = \min_{\langle \mathbf{x},\mathbf{y}\rangle=0} \angle(\mathbf{A}\mathbf{x}, \mathbf{A}\mathbf{y}). \tag{A.3}$$

*[Proof]:* The assertion can be shown by directly following the proof in [25, pp. 441-442]. □

*[Proof of Lemma 2.1]:* Let $T_{uv} = T_u \cup T_v$. Since $\boldsymbol{\Phi}\mathbf{u} = \boldsymbol{\Phi}_{T_{uv}}\mathbf{u}_{T_{uv}}$ and $\boldsymbol{\Phi}\mathbf{v} = \boldsymbol{\Phi}_{T_{uv}}\mathbf{v}_{T_{uv}}$, we have

$$\left|\cos\angle(\boldsymbol{\Phi}\mathbf{u}, \boldsymbol{\Phi}\mathbf{v})\right| = \left|\cos\angle\left(\boldsymbol{\Phi}_{T_{uv}}\mathbf{u}_{T_{uv}}, \boldsymbol{\Phi}_{T_{uv}}\mathbf{v}_{T_{uv}}\right)\right|. \tag{A.4}$$

Since $\mathbf{u}$ and $\mathbf{v}$, supported on, respectively, $T_u$ and $T_v$, are orthogonal, so are the pairs $(\mathbf{u}_{T_{uv}}, \mathbf{v}_{T_{uv}})$ and $(-\mathbf{u}_{T_{uv}}, \mathbf{v}_{T_{uv}})$. Also, since $|T_u \cup T_v| \leq K$ and arbitrary $K$ columns of $\boldsymbol{\Phi}$ are linearly independent, $\boldsymbol{\Phi}_{T_{uv}}$ is of full column rank. By Lemma A.2, it follows immediately that

$$\angle(\pm\boldsymbol{\Phi}_{T_{uv}}\mathbf{u}_{T_{uv}}, \boldsymbol{\Phi}_{T_{uv}}\mathbf{v}_{T_{uv}}) \geq \theta(\boldsymbol{\Phi}_{T_{uv}}). \tag{A.5}$$

From (A.5), we also have

$$\angle(\mp\boldsymbol{\Phi}_{T_{uv}}\mathbf{u}_{T_{uv}}, \boldsymbol{\Phi}_{T_{uv}}\mathbf{v}_{T_{uv}}) \leq \pi - \theta(\boldsymbol{\Phi}_{T_{uv}}). \tag{A.6}$$

With (A.5) and (A.6), it then follows that

$$\left|\cos\angle\left(\boldsymbol{\Phi}_{T_{uv}}\mathbf{u}_{T_{uv}}, \boldsymbol{\Phi}_{T_{uv}}\mathbf{v}_{T_{uv}}\right)\right| \leq \cos(\theta(\boldsymbol{\Phi}_{T_{uv}})). \tag{A.7}$$

It thus remains to show

$$\cos(\theta(\boldsymbol{\Phi}_{T_{uv}})) \leq \delta_K. \tag{A.8}$$

The assertion of Lemma 2.1 then follows from (A.4), (A.7) and (A.8). To prove (A.8), it is noted that, since $\kappa(\boldsymbol{\Phi}_{T_{uv}}) = \cot(\theta(\boldsymbol{\Phi}_{T_{uv}})/2)$ (cf. (A.2)), we have

$$\cos(\theta(\boldsymbol{\Phi}_{T_{uv}})) = \frac{\cot^2\left(\theta(\boldsymbol{\Phi}_{T_{uv}})/2\right) - 1}{\cot^2\left(\theta(\boldsymbol{\Phi}_{T_{uv}})/2\right) + 1} = \frac{\kappa^2(\boldsymbol{\Phi}_{T_{uv}}) - 1}{\kappa^2(\boldsymbol{\Phi}_{T_{uv}}) + 1} = 1 - \frac{2}{\kappa^2(\boldsymbol{\Phi}_{T_{uv}}) + 1}. \tag{A.9}$$

By definition, $\kappa(\boldsymbol{\Phi}_{T_{uv}}) = \overline{\sigma}(\boldsymbol{\Phi}_{T_{uv}})/\underline{\sigma}(\boldsymbol{\Phi}_{T_{uv}})$, where $\overline{\sigma}(\boldsymbol{\Phi}_{T_{uv}})$ and $\overline{\sigma}(\boldsymbol{\Phi}_{T_{uv}})$ are, respectively, the maximal and minimal singular values of $\boldsymbol{\Phi}_{T_{uv}}$. Hence, with Lemma A.1 it follows that

$$\kappa(\mathbf{\Phi}_{T_{uv}}) = \frac{\overline{\sigma}(\mathbf{\Phi}_{T_{uv}})}{\underline{\sigma}(\mathbf{\Phi}_{T_{uv}})} \leq \frac{\sqrt{1+\delta_K}}{\underline{\sigma}(\mathbf{\Phi}_{T_{uv}})} \leq \frac{\sqrt{1+\delta_K}}{\sqrt{1-\delta_K}}. \tag{A.10}$$

Inequality (A.8) can be directly obtained from (A.9) and (A.10). □

## B. Proof of Theorem 3.2

To prove Theorem 3.2, we need the following lemma.

**Lemma A.3:** Let $\mathbf{x}$ be a $K$-sparse vector with support $T$. Assume that the sensing matrix $\mathbf{\Phi} \in \mathbb{R}^{M \times N}$ satisfies RIP of order $K+1$ with RIC $\delta_{K+1}$. Then $\|\mathbf{\Phi}_T^* \mathbf{\Phi} \mathbf{x}\|_2 \geq \sqrt{1-\delta_{K+1}} \|\mathbf{\Phi} \mathbf{x}\|_2 \geq (1-\delta_{K+1}) \|\mathbf{x}\|_2$, or equivalently, $\|\mathbf{\Phi}_T^* \mathbf{\Phi}_T \mathbf{x}_T\|_2 \geq \sqrt{1-\delta_{K+1}} \|\mathbf{\Phi}_T \mathbf{x}_T\|_2 \geq (1-\delta_{K+1}) \|\mathbf{x}_T\|_2$.

[Proof]: Let $\mathbf{\Phi}_T = \mathbf{U}\mathbf{\Sigma}\mathbf{V}^*$ be an SVD of $\mathbf{\Phi}_T \in \mathbb{R}^{M \times |T|}$, where $\mathbf{U} \in \mathbb{R}^{M \times M}$ and $\mathbf{V} \in \mathbb{R}^{|T| \times |T|}$ are orthogonal, and $\mathbf{\Sigma} = \begin{bmatrix} \mathbf{\Sigma}_1 \\ \mathbf{0} \end{bmatrix} \in \mathbb{R}^{M \times |T|}$ with $\mathbf{\Sigma}_1 \in \mathbb{R}^{|T| \times |T|}$ being diagonal with positive diagonal entries.

The assertion of Lemma A.3 follows from the following set of relations:

$$\begin{aligned}
\|\mathbf{\Phi}_T^* \mathbf{\Phi} \mathbf{x}\|_2 &= \|\mathbf{\Phi}_T^* \mathbf{\Phi}_T \mathbf{x}_T\|_2 = \|\mathbf{V}\mathbf{\Sigma}^* \mathbf{U}^* \mathbf{U}\mathbf{\Sigma}\mathbf{V}^* \mathbf{x}_T\|_2 \\
&\overset{(a)}{=} \left\| \mathbf{V}[\mathbf{\Sigma}_1^* \ \mathbf{0}] \begin{bmatrix} \mathbf{U}_1^* \\ \mathbf{U}_2^* \end{bmatrix} [\mathbf{U}_1 \ \mathbf{U}_2] \begin{bmatrix} \mathbf{\Sigma}_1 \\ \mathbf{0} \end{bmatrix} \mathbf{V}^* \mathbf{x}_T \right\|_2 = \left\| \mathbf{V}[\mathbf{\Sigma}_1^* \ \mathbf{0}] \begin{bmatrix} \mathbf{U}_1^* \\ \mathbf{U}_2^* \end{bmatrix} [\mathbf{U}_1 \ \mathbf{0}] \begin{bmatrix} \mathbf{\Sigma}_1 \\ \mathbf{0} \end{bmatrix} \mathbf{V}^* \mathbf{x}_T \right\|_2 \\
&\overset{(b)}{=} \left\| [\mathbf{\Sigma}_1^* \ \mathbf{0}] \begin{bmatrix} \mathbf{U}_1^* \\ \mathbf{U}_2^* \end{bmatrix} [\mathbf{U}_1 \ \mathbf{0}] \begin{bmatrix} \mathbf{\Sigma}_1 \\ \mathbf{0} \end{bmatrix} \mathbf{V}^* \mathbf{x}_T \right\|_2 = \left\| \mathbf{\Sigma}_1^* \mathbf{U}_1^* [\mathbf{U}_1 \ \mathbf{0}] \begin{bmatrix} \mathbf{\Sigma}_1 \\ \mathbf{0} \end{bmatrix} \mathbf{V}^* \mathbf{x}_T \right\|_2 \\
&\overset{(c)}{\geq} \sqrt{1-\delta_{K+1}} \left\| \mathbf{U}_1^* [\mathbf{U}_1 \ \mathbf{0}] \begin{bmatrix} \mathbf{\Sigma}_1 \\ \mathbf{0} \end{bmatrix} \mathbf{V}^* \mathbf{x}_T \right\|_2 = \sqrt{1-\delta_{K+1}} \left\| \begin{bmatrix} \mathbf{U}_1^* \\ \mathbf{0} \end{bmatrix} [\mathbf{U}_1 \ \mathbf{0}] \begin{bmatrix} \mathbf{\Sigma}_1 \\ \mathbf{0} \end{bmatrix} \mathbf{V}^* \mathbf{x}_T \right\|_2 \\
&\overset{(d)}{=} \sqrt{1-\delta_{K+1}} \left\| \begin{bmatrix} \mathbf{U}_1^* \\ \mathbf{U}_2^* \end{bmatrix} [\mathbf{U}_1 \ \mathbf{0}] \begin{bmatrix} \mathbf{\Sigma}_1 \\ \mathbf{0} \end{bmatrix} \mathbf{V}^* \mathbf{x}_T \right\|_2 \\
&\overset{(e)}{=} \sqrt{1-\delta_{K+1}} \left\| [\mathbf{U}_1 \ \mathbf{0}] \begin{bmatrix} \mathbf{\Sigma}_1 \\ \mathbf{0} \end{bmatrix} \mathbf{V}^* \mathbf{x}_T \right\|_2 = \sqrt{1-\delta_{K+1}} \left\| [\mathbf{U}_1 \ \mathbf{U}_2] \begin{bmatrix} \mathbf{\Sigma}_1 \\ \mathbf{0} \end{bmatrix} \mathbf{V}^* \mathbf{x}_T \right\|_2 \\
&= \sqrt{1-\delta_{K+1}} \|\mathbf{U}\mathbf{\Sigma}\mathbf{V}^* \mathbf{x}_T\|_2 = \sqrt{1-\delta_{K+1}} \|\mathbf{\Phi}_T \mathbf{x}_T\|_2 = \sqrt{1-\delta_{K+1}} \|\mathbf{\Phi} \mathbf{x}\|_2 \\
&\overset{(f)}{\geq} (1-\delta_{K+1}) \|\mathbf{x}\|_2,
\end{aligned}$$

where in (a) we partition $\mathbf{U}$ into $\mathbf{U} = [\mathbf{U}_1 \ \mathbf{U}_2]$, where $\mathbf{U}_1 \in \mathbb{R}^{M \times |T|}$ and $\mathbf{U}_2 \in \mathbb{R}^{M \times (M-|T|)}$, (b) holds since $\mathbf{V}$ is orthogonal, (c) is true since Lemma A.1 asserts that the singular values of $\mathbf{\Phi}_T$ (appearing as the diagonal entries of $\mathbf{\Sigma}_1^*$) are no less than $\sqrt{1-\delta_{K+1}}$, (d) is true because $\mathbf{U}_2^* \mathbf{U}_1 = \mathbf{0}$, and (e) follows since $[\mathbf{U}_1 \ \mathbf{U}_2]^* = \mathbf{U}^*$ is orthogonal, and (f) is due to RIP. This thus proves the lemma. □

*Proof of Case (1):* The idea behind the proof is similar to that of Theorem 2.2. Recall that, in the *j*-th iteration, the index $\rho^j$ determined yields the maximal $\left|\langle \mathbf{\Phi e}_i, \mathbf{r}^{j-1}\rangle\right|$ (see Step 3.2 of Table I). We first claim that, if $\delta_{K+1}$ satisfies (2.2) and the threshold requirement (3.3) holds, such $\rho^j$'s, where $j = 1, \cdots, K$, belong to the support *T*. Also, according to the orthogonality property of OMP [8, Lemma 7], all the selected indexes $\rho^j$'s, $j = 1, \cdots, K$, are distinct. We will then prove that OMP with the stopping criterion $\|\mathbf{r}^j\|_2 \leq \varepsilon_1$ halts exactly after *K* iterations. The assertion of the theorem then follows.

To prove the claim, it suffices to show that the following two conditions hold for all $j = 1, \cdots, K$: in the *j*-th iteration, there exists some *K*-sparse $\mathbf{z}^j \in \mathbb{R}^N$ with support *T* such that $\mathbf{r}^{j-1}$ in Step 3.2 reads $\mathbf{r}^{j-1} = \mathbf{\Phi z}^j + \mathbf{w}$,

$$\left|\langle \mathbf{\Phi e}_i, \mathbf{r}^{j-1}\rangle\right| \leq \|\mathbf{\Phi z}^j\|_2 \delta_{K+1} + \varepsilon_1 \quad \text{for all} \quad i \notin T, \tag{A.11}$$

and

$$\left|\langle \mathbf{\Phi e}_i, \mathbf{r}^{j-1}\rangle\right| > \|\mathbf{\Phi z}^j\|_2 \delta_{K+1} + \varepsilon_1 \quad \text{for some} \quad i \in T. \tag{A.12}$$

Hence, we have $\rho^j \in T$ for all $j = 1, \cdots, K$. The proof of (A.11) and (A.12) is done by induction. In the first iteration $(j = 1)$, $\mathbf{r}^{j-1}$ needed to compute the inner product in Step 3.2 is

$$\mathbf{r}^0 = \mathbf{y} = \mathbf{\Phi x} + \mathbf{w}, \tag{A.13}$$

and hence

$$\left|\langle \mathbf{\Phi e}_i, \mathbf{r}^{j-1}\rangle\right| = \left|\langle \mathbf{\Phi e}_i, \mathbf{r}^0\rangle\right| = \left|\langle \mathbf{\Phi e}_i, \mathbf{\Phi x}\rangle + \langle \mathbf{\Phi e}_i, \mathbf{w}\rangle\right| \leq \left|\langle \mathbf{\Phi e}_i, \mathbf{\Phi x}\rangle\right| + \left|\langle \mathbf{\Phi e}_i, \mathbf{w}\rangle\right|. \tag{A.14}$$

Note that, for $i \notin T$, we have $\langle \mathbf{e}_i, \mathbf{x}\rangle = 0$ and $|\{i\} \cup T| = K + 1$. From Lemma 2.1 and since each column of $\mathbf{\Phi}$ is of unit-norm, it follows immediately that

$$\left|\langle \mathbf{\Phi e}_i, \mathbf{\Phi x}\rangle\right| \leq \|\mathbf{\Phi e}_i\|_2 \|\mathbf{\Phi x}\|_2 \left|\cos(\mathbf{\Phi e}_i, \mathbf{\Phi x})\right| \leq \|\mathbf{\Phi e}_i\|_2 \|\mathbf{\Phi x}\|_2 \delta_{K+1} = \|\mathbf{\Phi x}\|_2 \delta_{K+1} \quad \text{for all} \quad i \notin T. \tag{A.15}$$

Also, since $\|\mathbf{w}\|_2 \leq \varepsilon_1$, we have

$$\left|\langle \mathbf{\Phi e}_i, \mathbf{w}\rangle\right| \leq \|\mathbf{\Phi e}_i\|_2 \|\mathbf{w}\|_2 \leq \varepsilon_1. \tag{A.16}$$

From (A.14), (A.15) and (A.16), we have

$$\left|\langle \mathbf{\Phi e}_i, \mathbf{r}^0\rangle\right| \leq \|\mathbf{\Phi x}\|_2 \delta_{K+1} + \varepsilon_1 \quad \text{for all} \quad i \notin T, \tag{A.17}$$

i.e., (A.11) holds when $j = 1$ with $\mathbf{z}^1 = \mathbf{x}$. We then go on to show by contradiction that (A.12) is also true when $j = 1$ with $\mathbf{z}^1 = \mathbf{x}$. Assume otherwise that

$$\left|\langle \mathbf{\Phi e}_i, \mathbf{r}^0\rangle\right| \leq \|\mathbf{\Phi x}\|_2 \delta_{K+1} + \varepsilon_1 \quad \textit{for all} \quad i \in T. \tag{A.18}$$

Then, it follows from (A.18) that

$$\left\|\mathbf{\Phi}_T^* \mathbf{r}^0\right\|_2 = \sqrt{\sum_{i \in T} \left|\langle \mathbf{\Phi e}_i, \mathbf{r}^0 \rangle\right|^2} \leq \sqrt{K}\left(\|\mathbf{\Phi x}\|_2 \delta_{K+1} + \varepsilon_1\right) = \sqrt{K}\delta_{K+1}\|\mathbf{\Phi x}\|_2 + \sqrt{K}\varepsilon_1. \quad (A.19)$$

Also,

$$\left\|\mathbf{\Phi}_T^* \mathbf{r}^0\right\|_2 = \left\|\mathbf{\Phi}_T^*(\mathbf{\Phi x} + \mathbf{w})\right\|_2 \geq \left\|\mathbf{\Phi}_T^* \mathbf{\Phi x}\right\|_2 - \left\|\mathbf{\Phi}_T^* \mathbf{w}\right\|_2 \overset{(a)}{\geq} \sqrt{1-\delta_{K+1}}\|\mathbf{\Phi x}\|_2 - \left\|\mathbf{\Phi}_T^* \mathbf{w}\right\|_2 \overset{(b)}{\geq} \sqrt{1-\delta_{K+1}}\|\mathbf{\Phi x}\|_2 - \sqrt{1+\delta_{K+1}}\varepsilon_1, \quad (A.20)$$

where (a) follows from Lemma A.3 and (b) is true due to Lemma A.1 and $\|\mathbf{w}\|_2 \leq \varepsilon_1$. Under assumption (3.4), it can be shown that

$$\sqrt{1-\delta_{K+1}}\|\mathbf{\Phi x}\|_2 - \sqrt{1+\delta_{K+1}}\varepsilon_1 > \sqrt{K}\delta_{K+1}\|\mathbf{\Phi x}\|_2 + \sqrt{K}\varepsilon_1. \quad (A.21)$$

Then, combining (A.20) and (A.21) yields

$$\left\|\mathbf{\Phi}_T^* \mathbf{r}^0\right\|_2 > \sqrt{K}\delta_{K+1}\|\mathbf{\Phi x}\|_2 + \sqrt{K}\varepsilon_1, \quad (A.22)$$

which contradicts with (A.19). This then implies (A.12) is true for $j=1$. To verify (A.21), we first write

$$\sqrt{1-\delta_{K+1}}\|\mathbf{\Phi x}\|_2 - \sqrt{1+\delta_{K+1}}\varepsilon_1$$
$$= \sqrt{K}\delta_{K+1}\|\mathbf{\Phi x}\|_2 + \sqrt{K}\varepsilon_1 + \underbrace{\left(\sqrt{1-\delta_{K+1}} - \sqrt{K}\delta_{K+1}\right)}_{\triangleq \gamma}\|\mathbf{\Phi x}\|_2 - \left(\sqrt{1+\delta_{K+1}} + \sqrt{K}\right)\varepsilon_1. \quad (A.23)$$

One can check that under the assumption $\delta_{K+1} < \dfrac{\sqrt{4K+1}-1}{2K}$, we have $\gamma \geq 0$. Thus, it follows from (A.23) that

$$\sqrt{1-\delta_{K+1}}\|\mathbf{\Phi x}\|_2 - \sqrt{1+\delta_{K+1}}\varepsilon_1$$
$$\overset{(a)}{\geq} \sqrt{K}\delta_{K+1}\|\mathbf{\Phi x}\|_2 + \sqrt{K}\varepsilon_1 + \underbrace{\left(\sqrt{1-\delta_{K+1}} - \sqrt{K}\delta_{K+1}\right)}_{=\gamma}\sqrt{1-\delta_{K+1}}\|\mathbf{x}\|_2 - \left(\sqrt{1+\delta_{K+1}} + \sqrt{K}\right)\varepsilon_1$$
$$\overset{(b)}{>} \sqrt{K}\delta_{K+1}\|\mathbf{\Phi x}\|_2 + \sqrt{K}\varepsilon_1$$
$$+ \left((1-\delta_{K+1}) - \sqrt{1-\delta_{K+1}}\sqrt{K}\delta_{K+1}\right)\sqrt{K}\dfrac{(\sqrt{1+\delta_{K+1}}+1)\varepsilon_1}{1-\delta_{K+1}-\sqrt{1-\delta_{K+1}}\sqrt{K}\delta_{K+1}} - \left(\sqrt{1+\delta_{K+1}} + \sqrt{K}\right)\varepsilon_1$$
$$\overset{(c)}{=} \sqrt{K}\delta_{K+1}\|\mathbf{\Phi x}\|_2 + \sqrt{K}\varepsilon_1 + \underbrace{\varepsilon_1(\sqrt{K}-1)\sqrt{1+\delta_{K+1}}}_{\geq 0}$$
$$\geq \sqrt{K}\delta_{K+1}\|\mathbf{\Phi x}\|_2 + \sqrt{K}\varepsilon_1, \quad (A.24)$$

where (a) follows since $\mathbf{\Phi}$ satisfies RIP and $\mathbf{x}$ is a $K$-sparse vector, (b) follows since $\mathbf{x}$ is $K$-sparse

(thus, with at most $K$ nonzero entries), and, due to assumption (3.4), and (c) holds after some straightforward manipulations.

We move on to the next iteration. Since the index selected in the first iteration belongs to $T$, in the second iteration ($j=2$), $\mathbf{r}^{j-1}$ needed in Step 3.2 is[2]

$$\mathbf{r}^{j-1} = \mathbf{r}^1 = \mathbf{y} - \mathbf{\Phi}_{\Omega^1}\mathbf{q}_{\Omega^1} = \mathbf{\Phi}\mathbf{x} - \mathbf{\Phi}_{\Omega^1}\mathbf{q}_{\Omega^1} + \mathbf{w} = \mathbf{\Phi}\mathbf{x} - \mathbf{\Phi}\tilde{\mathbf{q}}_{\Omega^1} + \mathbf{w} = \mathbf{\Phi}\mathbf{z}^2 + \mathbf{w}, \tag{A.25}$$

where $\tilde{\mathbf{q}}_{\Omega^1} \in \mathbb{R}^N$ is a 1-sparse vector with all entries equal to zero except the one indexed by $\Omega^1 \subset T$. Then $\mathbf{z}^2 = \mathbf{x} - \tilde{\mathbf{q}}_{\Omega^1}$ is a $K$-sparse vector with support $T$. By following essentially the same procedures as in the first iteration, (A.11) and (A.12) can be shown to be true for $j=2$ with the $K$-sparse vector $\mathbf{z}^2$. By repeating such procedures, one can inductively show that the first $K$ selected indexes all belong to the support $T$. Also, since the selected indexes are distinct, the columns of $\mathbf{\Phi}_{\Omega^K} \in \mathbb{R}^{M \times K}$ are those of $\mathbf{\Phi}$ indexed by $T$.

Now we turn to show that, with the stopping criterion $\|\mathbf{r}^j\|_2 \leq \varepsilon_1$, OMP stops exactly after $K$ iterations. This can be done by showing that

$$\|\mathbf{r}^j\|_2 > \varepsilon_1 \text{ for all } j = 0, \cdots, K-1, \text{ and } \|\mathbf{r}^j\|_2 \leq \varepsilon_1 \text{ for } j = K. \tag{A.26}$$

The first condition in (A.26) holds since, for $j = 0, \cdots, K-1$,

$$\begin{aligned}
\|\mathbf{r}^j\|_2 &= \|\mathbf{\Phi}\mathbf{x} - \mathbf{\Phi}_{\Omega^j}\mathbf{q}_{\Omega^j} + \mathbf{w}\|_2 = \|\mathbf{\Phi}\mathbf{x} - \mathbf{\Phi}\tilde{\mathbf{q}}_{\Omega^j} + \mathbf{w}\|_2 \geq \|\mathbf{\Phi}(\mathbf{x} - \tilde{\mathbf{q}}_{\Omega^j})\|_2 - \|\mathbf{w}\|_2 \geq \|\mathbf{\Phi}(\mathbf{x} - \tilde{\mathbf{q}}_{\Omega^j})\|_2 - \varepsilon_1 \\
&\stackrel{(a)}{\geq} \sqrt{1-\delta_{K+1}}\|\mathbf{x} - \tilde{\mathbf{q}}_{\Omega^j}\|_2 - \varepsilon_1 = \sqrt{1-\delta_{K+1}}\sqrt{\sum_{i=1}^N (\mathbf{x} - \tilde{\mathbf{q}}_{\Omega^j})_i^2} - \varepsilon_1 \\
&\geq \sqrt{1-\delta_{K+1}}\sqrt{\sum_{i \in T \setminus \Omega^j}(\mathbf{x} - \tilde{\mathbf{q}}_{\Omega^j})_i^2} - \varepsilon_1 = \sqrt{1-\delta_{K+1}}\sqrt{\sum_{i \in T \setminus \Omega^j}(\mathbf{x})_i^2} - \varepsilon_1 \\
&\stackrel{(b)}{\geq} \sqrt{1-\delta_{K+1}}\sqrt{K-j}\frac{(\sqrt{1+\delta_{K+1}}+1)\varepsilon_1}{1-\delta_{K+1}-\sqrt{1-\delta_{K+1}}\sqrt{K}\delta_{K+1}} - \varepsilon_1 \\
&\geq \sqrt{1-\delta_{K+1}}\frac{(\sqrt{1-\delta_{K+1}}+1)\varepsilon_1}{1-\delta_{K+1}-\sqrt{1-\delta_{K+1}}\sqrt{K}\delta_{K+1}} - \varepsilon_1 \\
&= \frac{(1-\delta_{K+1}) + \sqrt{1-\delta_{K+1}} - (1-\delta_{K+1} - \sqrt{1-\delta_{K+1}}\sqrt{K}\delta_{K+1})}{1-\delta_{K+1} - \sqrt{1-\delta_{K+1}}\sqrt{K}\delta_{K+1}}\varepsilon_1 \\
&= \frac{\sqrt{1-\delta_{K+1}} + \sqrt{1-\delta_{K+1}}\sqrt{K}\delta_{K+1}}{1-\delta_{K+1} - \sqrt{1-\delta_{K+1}}\sqrt{K}\delta_{K+1}}\varepsilon_1 \geq \frac{(1-\delta_{K+1}) + (1-\delta_{K+1})\sqrt{K}\delta_{K+1}}{(1-\delta_{K+1}) - \sqrt{1-\delta_{K+1}}\sqrt{K}\delta_{K+1}}\varepsilon_1 > \varepsilon_1, \tag{A.27}
\end{aligned}$$

---

[2]. We remind the readers of the following notation usage that is frequently adopted throughout the appendix: For $\mathbf{u} \in \mathbb{R}^N$, $\mathbf{u}_S \in \mathbb{R}^{|S|}$ denotes the vector whose entries consist of those of $\mathbf{u}$ indexed by the subset $S \subset \{1, \cdots, N\}$, and $\tilde{\mathbf{u}}_S \in \mathbb{R}^N$ is the zero-padded version of $\mathbf{u}_S$ such that $(\tilde{\mathbf{u}}_S)_i = (\mathbf{u})_i$ for $i \in S$ and $(\tilde{\mathbf{u}}_S)_i = 0$ otherwise (thus, $\tilde{\mathbf{u}}_S$ is $|S|$-sparse with support $S$).

where (a) follows since $\mathbf{x} - \tilde{\mathbf{q}}_{\Omega^j}$ is supported on $T$ (recalling the support of $\tilde{\mathbf{q}}_{\Omega^j}$ is $\Omega^j \subset T$) and using RIP, and (b) holds by assumption (3.4). The second condition in (A.26) is also true since, for $j = K$,

$$\left\|\mathbf{r}^K\right\|_2 = \left\|\mathbf{y} - \boldsymbol{\Phi}_{\Omega^K}\mathbf{q}_{\Omega^K}\right\|_2 \overset{(a)}{\leq} \left\|\mathbf{y} - \boldsymbol{\Phi}_{\Omega^K}\mathbf{x}_T\right\|_2 \overset{(b)}{=} \left\|\mathbf{y} - \boldsymbol{\Phi}_T\mathbf{x}_T\right\|_2 = \left\|\mathbf{y} - \boldsymbol{\Phi}\mathbf{x}\right\|_2 = \|\mathbf{w}\|_2 \leq \varepsilon_1, \quad \text{(A.28)}$$

where (a) follows since, by definition, $\mathbf{q}_{\Omega^K}$ yields the minimal squares errors at the $K$-th iteration (see Step 3.4 in Table I) and (b) is true because $\boldsymbol{\Phi}_{\Omega^K} = \boldsymbol{\Phi}_T$ as asserted previously.

*Proof of Case (2):* As in Case (1), we will first show that OMP with the stopping criterion $\left\|\boldsymbol{\Phi}^*\mathbf{r}^j\right\|_\infty = \max_{i=1,\ldots,N}\left|\left\langle\boldsymbol{\Phi}\mathbf{e}_i, \mathbf{r}^j\right\rangle\right| \leq \varepsilon_2$ can correctly identify the support in the first $K$ iterations. To show that the indexes selected in the first $K$ iterations all belong to $T$, it suffices to prove the following claim: in the $j$-th iteration, $j = 1,\ldots,K$, there exists some $K$-sparse $\mathbf{z}^j \in \mathbb{R}^N$ with support $T$ such that $\mathbf{r}^{j-1}$ in Step 3.2 reads $\mathbf{r}^{j-1} = \boldsymbol{\Phi}\mathbf{z}^j + \mathbf{w}$,

$$\left|\left\langle\boldsymbol{\Phi}\mathbf{e}_i, \mathbf{r}^{j-1}\right\rangle\right| \leq \left\|\boldsymbol{\Phi}\mathbf{z}^j\right\|_2 \delta_{K+1} + \varepsilon_2 \quad \text{for all} \quad i \notin T, \quad \text{(A.29)}$$

and

$$\left|\left\langle\boldsymbol{\Phi}\mathbf{e}_i, \mathbf{r}^{j-1}\right\rangle\right| > \left\|\boldsymbol{\Phi}\mathbf{z}^j\right\|_2 \delta_{K+1} + \varepsilon_2 \quad \text{for some} \quad i \in T. \quad \text{(A.30)}$$

That is, we have $\rho^j \in T$ for all $j = 1,\ldots,K$. Also, since the index, once identified, will not be selected again [8], the assertion of the theorem is thus proved. To show that (A.29) holds in the first iteration ($j = 1$), it is noted from (A.14) and (A.15) that

$$\left|\left\langle\boldsymbol{\Phi}\mathbf{e}_i, \mathbf{r}^{j-1}\right\rangle\right| = \left|\left\langle\boldsymbol{\Phi}\mathbf{e}_i, \mathbf{r}^0\right\rangle\right| \leq \|\boldsymbol{\Phi}\mathbf{x}\|_2 \delta_{K+1} + \left|\left\langle\boldsymbol{\Phi}\mathbf{e}_i, \mathbf{w}\right\rangle\right|, \text{ for all } i \notin T. \quad \text{(A.31)}$$

Since $\left\|\boldsymbol{\Phi}^*\mathbf{w}\right\|_\infty \leq \varepsilon_2$, we have

$$\left|\left\langle\boldsymbol{\Phi}\mathbf{e}_i, \mathbf{w}\right\rangle\right| \leq \varepsilon_2 \quad \text{for all} \quad i = 1,\ldots,N. \quad \text{(A.32)}$$

With (A.31) and (A.32), it is easy to see that (A.29) holds when $j = 1$ with $\mathbf{z}^1 = \mathbf{x}$. We turn to show by contradiction that (A.30) is also true for $j = 1$ with $\mathbf{z}^1 = \mathbf{x}$. Assume otherwise that

$$\left|\left\langle\boldsymbol{\Phi}\mathbf{e}_i, \mathbf{r}^{j-1}\right\rangle\right| = \left|\left\langle\boldsymbol{\Phi}\mathbf{e}_i, \mathbf{r}^0\right\rangle\right| \leq \left\|\boldsymbol{\Phi}\mathbf{z}^j\right\|_2 \delta_{K+1} + \varepsilon_2 = \|\boldsymbol{\Phi}\mathbf{x}\|_2 \delta_{K+1} + \varepsilon_2 \text{ for all } i \in T. \quad \text{(A.33)}$$

It follows immediately from (A.33) that

$$\left\|\boldsymbol{\Phi}_T^*\mathbf{r}^0\right\|_2 = \sqrt{\sum_{i \in T}\left|\left\langle\boldsymbol{\Phi}\mathbf{e}_i, \mathbf{r}^0\right\rangle\right|^2} \leq \sqrt{K}\left(\|\boldsymbol{\Phi}\mathbf{x}\|_2 \delta_{K+1} + \varepsilon_2\right) = \sqrt{K}\delta_{K+1}\|\boldsymbol{\Phi}\mathbf{x}\|_2 + \sqrt{K}\varepsilon_2. \quad \text{(A.34)}$$

Also,

$$\left\|\mathbf{\Phi}_T^*\mathbf{r}^0\right\|_2 = \left\|\mathbf{\Phi}_T^*(\mathbf{\Phi}\mathbf{x}+\mathbf{w})\right\|_2 \geq \left\|\mathbf{\Phi}_T^*\mathbf{\Phi}\mathbf{x}\right\|_2 - \left\|\mathbf{\Phi}_T^*\mathbf{w}\right\|_2 \overset{(a)}{\geq} \sqrt{1-\delta_{K+1}}\left\|\mathbf{\Phi}\mathbf{x}\right\|_2 - \left\|\mathbf{\Phi}_T^*\mathbf{w}\right\|_2$$
$$= \sqrt{1-\delta_{K+1}}\left\|\mathbf{\Phi}\mathbf{x}\right\|_2 - \sqrt{\sum_{i\in T}\left|\left\langle\mathbf{\Phi}\mathbf{e}_i,\mathbf{w}\right\rangle\right|^2} \overset{(b)}{\geq} \sqrt{1-\delta_{K+1}}\left\|\mathbf{\Phi}\mathbf{x}\right\|_2 - \sqrt{K}\varepsilon_2, \quad (A.35)$$

where (a) follows according to Lemma A.3 and (b) holds on account of (A.32). With (3.5), one can verify that

$$\sqrt{1-\delta_{K+1}}\left\|\mathbf{\Phi}\mathbf{x}\right\|_2 - \sqrt{K}\varepsilon_2 > \sqrt{K}\delta_{K+1}\left\|\mathbf{\Phi}\mathbf{x}\right\|_2 + \sqrt{K}\varepsilon_2. \quad (A.36)$$

Then, combining (A.35) and (A.36) yields

$$\left\|\mathbf{\Phi}_T^*\mathbf{r}^0\right\|_2 > \sqrt{K}\delta_{K+1}\left\|\mathbf{\Phi}\mathbf{x}\right\|_2 + \sqrt{K}\varepsilon_1, \quad (A.37)$$

which contradicts with (A.34). This then implies (A.30) is true for $j=1$. Now we turn to show that (A.36) is true. Since

$$\left\|\mathbf{\Phi}\mathbf{x}\right\|_2 \overset{(a)}{\geq} \sqrt{1-\delta_{K+1}}\left\|\mathbf{x}\right\|_2 = \sqrt{1-\delta_{K+1}}\left\|\mathbf{x}_T\right\|_2 \overset{(b)}{>} \frac{\sqrt{K}(\sqrt{K}+1)\varepsilon_2}{\sqrt{1-\delta_{K+1}}-\sqrt{K}\delta_{K+1}} \geq \frac{2\sqrt{K}\varepsilon_2}{\sqrt{1-\delta_{K+1}}-\sqrt{K}\delta_{K+1}}, \quad (A.38)$$

where (a) is due to RIP and (b) follows with assumption (3.5), it can be directly verified that (A.36) holds. By following essentially the same procedures as in Case (1), one can verify that, under the threshold requirement (3.5), both (A.29) and (A.30) remain true for $j=2,\ldots,K$. Thus, the OMP algorithm will correctly identify the support $T$ in the first $K$ iterations.

Then, we show that OMP with the stopping criterion $\max_{i=1,\ldots,N}\left|\left\langle\mathbf{\Phi}\mathbf{e}_i,\mathbf{r}^j\right\rangle\right|\leq\varepsilon_2$ will not halt during the first $K$ iterations. This can be done by proving that

$$\max_{i=1,\ldots,N}\left|\left\langle\mathbf{\Phi}\mathbf{e}_i,\mathbf{r}^j\right\rangle\right| > \varepsilon_2 \text{ for all } j=0,\cdots,K-1. \quad (A.39)$$

To prove (A.39), assume otherwise that

$$\max_{i=1,\ldots,N}\left|\left\langle\mathbf{\Phi}\mathbf{e}_i,\mathbf{r}^j\right\rangle\right| \leq \varepsilon_2 \text{ for some } 0\leq j_0\leq K-1. \quad (A.40)$$

Then, associated with such $j_0$ it can be deduced by using (A.40) that

$$\left\|\mathbf{\Phi}_{T\setminus\mathbf{\Omega}^{j_0}}^*\mathbf{r}^{j_0}\right\|_2 = \sqrt{\sum_{i\in T\setminus\mathbf{\Omega}^{j_0}}\left|\left\langle\mathbf{\Phi}\mathbf{e}_i,\mathbf{r}^{j_0}\right\rangle\right|^2} \leq \sqrt{K-j_0}\varepsilon_2. \quad (A.41)$$

Also, we see that, for all $j=0,\cdots,K-1$,

$$\left\|\mathbf{\Phi}_{T\setminus\mathbf{\Omega}^j}^*\mathbf{r}^j\right\|_2 \stackrel{(a)}{=} \left\|\mathbf{\Phi}_T^*\mathbf{r}^j\right\|_2$$

$$\stackrel{(b)}{=} \left\|\mathbf{\Phi}_T^*\mathbf{\Phi}(\mathbf{x}-\tilde{\mathbf{q}}_{\mathbf{\Omega}^j}) + \mathbf{\Phi}_T^*\mathbf{w}\right\|_2 \geq \left\|\mathbf{\Phi}_T^*\mathbf{\Phi}(\mathbf{x}-\tilde{\mathbf{q}}_{\mathbf{\Omega}^j})\right\|_2 - \left\|\mathbf{\Phi}_T^*\mathbf{w}\right\|_2$$

$$= \left\|\mathbf{\Phi}_T^*\mathbf{\Phi}(\mathbf{x}-\tilde{\mathbf{q}}_{\mathbf{\Omega}^j})\right\|_2 - \sqrt{\sum_{i\in T}|\langle\mathbf{\Phi}\mathbf{e}_i,\mathbf{w}\rangle|^2}$$

$$\stackrel{(c)}{\geq} \left\|\mathbf{\Phi}_T^*\mathbf{\Phi}(\mathbf{x}-\tilde{\mathbf{q}}_{\mathbf{\Omega}^j})\right\|_2 - \sqrt{K}\varepsilon_2$$

$$\stackrel{(d)}{\geq} (1-\delta_{K+1})\left\|\mathbf{x}-\tilde{\mathbf{q}}_{\mathbf{\Omega}^j}\right\|_2 - \sqrt{K}\varepsilon_2 = (1-\delta_{K+1})\sqrt{\sum_{i\in T}(\mathbf{x}-\tilde{\mathbf{q}}_{\mathbf{\Omega}^j})_i^2} - \sqrt{K}\varepsilon_2$$

$$\geq (1-\delta_{K+1})\sqrt{\sum_{i\in T\setminus\mathbf{\Omega}^j}(\mathbf{x}-\tilde{\mathbf{q}}_{\mathbf{\Omega}^j})_i^2} - \sqrt{K}\varepsilon_2$$

$$= (1-\delta_{K+1})\sqrt{\sum_{i\in T\setminus\mathbf{\Omega}^j}(\mathbf{x})_i^2} - \sqrt{K}\varepsilon_2$$

$$\stackrel{(e)}{>} (1-\delta_{K+1})\sqrt{K-j}\frac{(\sqrt{K}+1)\varepsilon_1}{1-\delta_{K+1}-\sqrt{1-\delta_{K+1}}\sqrt{K}\delta_{K+1}} - \sqrt{K}\varepsilon_2$$

$$\geq (1-\delta_{K+1})\sqrt{K-j}\frac{(\sqrt{K}+1)\varepsilon_1}{1-\delta_{K+1}} - \sqrt{K}\varepsilon_2 = \sqrt{K-j}\left((\sqrt{K}+1)-\sqrt{K}/\sqrt{K-j}\right)\varepsilon_2$$

$$\geq \sqrt{K-j}(\sqrt{K}+1-\sqrt{K})\varepsilon_2 \geq \sqrt{K-j}\varepsilon_2, \tag{A.42}$$

where (a) follows since $\mathbf{r}^j$ is orthogonal to the columns of $\mathbf{\Phi}$ indexed by the selected indexes in $\mathbf{\Omega}^j$, i.e. $(\mathbf{\Phi}\mathbf{e}_i)^*\mathbf{r}^{j-1}=0$ for all $i\in\mathbf{\Omega}^j$, (b) follows since $\mathbf{r}^j$ in Step 3.8 reads $\mathbf{r}^j = \mathbf{\Phi}\mathbf{x}+\mathbf{w}-\mathbf{\Phi}_{\mathbf{\Omega}^j}\mathbf{q}_{\mathbf{\Omega}^j} = \mathbf{\Phi}\mathbf{x}+\mathbf{w}-\mathbf{\Phi}\tilde{\mathbf{q}}_{\mathbf{\Omega}^j}$, (c) is true due to the assumption $\|\mathbf{\Phi}^*\mathbf{w}\|_\infty \leq \varepsilon_2$, (d) holds with the fact that $\mathbf{x}-\tilde{\mathbf{q}}_{\mathbf{\Omega}^j}$ is supported on $T$ (since the support of $\tilde{\mathbf{q}}_{\mathbf{\Omega}^j}$ is $\mathbf{\Omega}^j \subset T$) and by using Lemma A.3, and (e) is obtained by using the assumption (3.5). Since (A.42) contradicts with (A.41), the assertion (A.39) is true. □

## C. Proof of Theorem 4.3

In the sequel, for $S\subset\{1,\cdots,N\}$ with $|S|<K$, $\mathbf{P}_S \triangleq \mathbf{\Phi}_S(\mathbf{\Phi}_S^*\mathbf{\Phi}_S)^{-1}\mathbf{\Phi}_S^*$ represents the orthogonal projection onto $\mathcal{R}(\mathbf{\Phi}_S)$. We need the next lemma to complete the proof.

**Lemma A.4:** Assume that $\mathbf{\Phi}$ satisfies RIP of order $K$ with RIC $\delta_K$. Let $S\subset\{1,\cdots,N\}$ with $|S|<K$, and $\mathbf{P}_S$ be the orthogonal projection onto $\mathcal{R}(\mathbf{\Phi}_S)$. Then for all $(K-|S|)$-sparse $\mathbf{x}$ whose support does not overlap with $S$, we have

$$\left\|\mathbf{P}_S\mathbf{\Phi}\mathbf{x}\right\|_2^2 \leq \delta_K^2\left\|\mathbf{\Phi}\mathbf{x}\right\|_2^2 \leq (1+\delta_K)\delta_K^2\left\|\mathbf{x}\right\|_2^2, \tag{A.43}$$

and

$$\left\|(\mathbf{I}-\mathbf{P}_S)\mathbf{\Phi}\mathbf{x}\right\|_2^2 \geq (1-\delta_K^2)\left\|\mathbf{\Phi}\mathbf{x}\right\|_2^2 \geq (1-\delta_K)(1-\delta_K^2)\left\|\mathbf{x}\right\|_2^2. \tag{A.44}$$

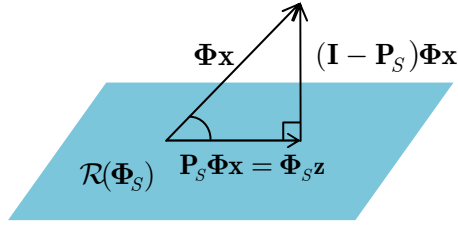

Fig. 3. Schematic description of orthogonal projection of $\mathbf{\Phi x}$ onto the column space of $\mathbf{\Phi}_S$.

*[Proof of Lemma A.4]:* Since $\mathbf{P}_S = \mathbf{\Phi}_S(\mathbf{\Phi}_S^*\mathbf{\Phi}_S)^{-1}\mathbf{\Phi}_S^*$, we have

$$\mathbf{P}_S\mathbf{\Phi x} = \mathbf{\Phi}_S(\mathbf{\Phi}_S^*\mathbf{\Phi}_S)^{-1}\mathbf{\Phi}_S^*\mathbf{\Phi x} = \mathbf{\Phi}_S\mathbf{z}, \text{ with } \mathbf{z} = (\mathbf{\Phi}_S^*\mathbf{\Phi}_S)^{-1}\mathbf{\Phi}_S^*\mathbf{\Phi x}. \tag{A.45}$$

It follows that (see Figure 3)

$$\|\mathbf{P}_S\mathbf{\Phi x}\|_2^2 = \|\mathbf{\Phi x}\|_2^2 \left|\cos\angle(\mathbf{\Phi x}, \mathbf{\Phi}_S\mathbf{z})\right|^2 \text{ and } \|(\mathbf{I}-\mathbf{P}_S)\mathbf{\Phi x}\|_2^2 = \|\mathbf{\Phi x}\|_2^2 \left|\sin\angle(\mathbf{\Phi x}, \mathbf{\Phi}_S\mathbf{z})\right|^2. \tag{A.46}$$

Since $\mathbf{\Phi}_S\mathbf{z}$ can be written as

$$\mathbf{\Phi}_S\mathbf{z} = \mathbf{\Phi}\tilde{\mathbf{z}}, \tag{A.47}$$

where $\tilde{\mathbf{z}}$ is obtained by padding zeros to $\mathbf{z}$ (the support of $\tilde{\mathbf{z}}$ is $S$), we obtain

$$\angle(\mathbf{\Phi x}, \mathbf{\Phi}_S\mathbf{z}) = \angle(\mathbf{\Phi x}, \mathbf{\Phi}\tilde{\mathbf{z}}). \tag{A.48}$$

Note that $\mathbf{x}$ and $\tilde{\mathbf{z}}$ are orthogonal since the supports of $\mathbf{x}$ and $\tilde{\mathbf{z}}$ do not overlap. According to Lemma 2.1, it follows

$$\left|\cos\angle(\mathbf{\Phi x}, \mathbf{\Phi}\tilde{\mathbf{z}})\right|^2 \leq \delta_K^2 \tag{A.49}$$

and thus

$$\left|\sin\angle(\mathbf{\Phi x}, \mathbf{\Phi}\tilde{\mathbf{z}})\right|^2 = 1 - \left|\cos\angle(\mathbf{\Phi x}, \mathbf{\Phi}\tilde{\mathbf{z}})\right|^2 \geq 1 - \delta_K^2 \quad . \tag{A.50}$$

Combining (A.46), (A.49) and (A.50) yields $\|\mathbf{P}_S\mathbf{\Phi x}\|_2^2 \leq \delta_K^2\|\mathbf{\Phi x}\|_2^2$ and $\|(\mathbf{I}-\mathbf{P}_S)\mathbf{\Phi x}\|_2^2 > (1-\delta_K^2)\|\mathbf{\Phi x}\|_2^2$. Also, since $\mathbf{\Phi}$ satisfies RIP, (A.43) and (A.44) directly follow. □

*[Proof of Theorem 4.3]:* By definition, $\mathbf{Q}$ is the orthogonal projection onto the orthogonal complement of $\mathcal{R}(\mathbf{\Phi}_{T_d})$. Under the assumptions that $\mathbf{\Phi}$ satisfies the RIP of order $K$ with RIC $\delta_K$, and that the support of the $(K-|T_d|)$-sparse $\mathbf{x}$ does not overlap with $T_d$, it then follows from Lemma A.4 that

$$\|\mathbf{Q}\mathbf{\Phi x}\|_2^2 \geq (1-\delta_K)(1-\delta_K^2)\|\mathbf{x}\|_2^2. \tag{A.51}$$

Also, we have

$$\|\mathbf{Q}\mathbf{\Phi x}\|_2^2 \leq \|\mathbf{\Phi x}\|_2^2 \leq (1+\delta_K)\|\mathbf{x}\|_2^2. \tag{A.52}$$

Combining (A.51) with (A.52) immediately gives

$$(1-\delta_K)(1-\delta_K^2)\|\mathbf{x}\|_2^2 \leq \|\mathbf{Q}\mathbf{\Phi}\mathbf{x}\|_2^2 \leq (1+\delta_K)\|\mathbf{x}\|_2^2. \qquad (A.53)$$

The assertion thus follows from (A.53). □

*D. Proof of Theorem 4.5*

It suffices to show

$$\|\mathbf{r}^j\|_2 < \|\mathbf{r}^{j-1}\|_2 \text{ if } \|\mathbf{r}^{j-1}\|_2 > 0, \qquad (A.54)$$

that is, the norm of the residual is reduced iteration by iteration provided that $\|\mathbf{r}^{j-1}\|_2 > 0$. If (A.54) is true, the stopping criterion $\|\mathbf{r}^j\|_2 \geq \|\mathbf{r}^{j-1}\|_2$ will be satisfied only when $\|\mathbf{r}^{j_0-1}\|_2 = 0$ for some $j_0 - 1$[3]. Indeed, if $\|\mathbf{r}^{j_0-1}\|_2 = 0$, SP will halt at the $j_0$-th iteration since $\|\mathbf{r}^{j_0}\|_2 \geq \|\mathbf{r}^{j_0-1}\|_2 = 0$. It then remains to show that, for such a $j_0$, we also have $\|\mathbf{r}^{j_0}\|_2 = 0$, implying that the estimated sparse vector after $j_0$ iterations is exactly $\mathbf{x}$. Hence, the proof of theorem 4.5 is completed. To prove (A.54), we need the next two lemmas.

*Lemma A.5 [28]:* Assume that the sensing matrix $\mathbf{\Phi}$ satisfies RIP of order $K$ with RIC $\delta_K$. If $T_1 \cap T_2 = \phi$ and $|T_1 \cup T_2| \leq K$, then we have

$$\|\mathbf{\Phi}_{T_1}^* \mathbf{\Phi}_{T_2} \mathbf{x}_{T_2}\|_2 \leq \delta_K \|\mathbf{x}_{T_2}\|_2. \qquad (A.55)$$

□

*Lemma A.6:* Assume that the sensing matrix $\mathbf{\Phi}$ satisfies RIP of order $3K$ with RIC $\delta_{3K}$. Let $\mathbf{\Omega}^j$ be the estimated support in the *j*-th iteration of the SP algorithm. Then we have

$$\|\mathbf{x}_{T \setminus \mathbf{\Omega}^j}\|_2 \leq \alpha \|\mathbf{x}_{T \setminus \mathbf{\Omega}^{j-1}}\|_2, \qquad (A.56)$$

where $\alpha$ is defined in (4.17).
*[Proof]:* The proof is relegated to the end of this appendix. □

To claim (A.54), we first recall from Steps 3.7-3.8 in Table II that the residual vector is

$$\mathbf{r}^j = \mathbf{y} - \mathbf{\Phi}_{\mathbf{\Omega}^j}\mathbf{q}_{\mathbf{\Omega}^j} = \left[\mathbf{I} - \mathbf{\Phi}_{\mathbf{\Omega}^j}(\mathbf{\Phi}_{\mathbf{\Omega}^j}^*\mathbf{\Phi}_{\mathbf{\Omega}^j})^{-1}\mathbf{\Phi}_{\mathbf{\Omega}^j}^*\right]\mathbf{y} = [\mathbf{I} - \mathbf{P}_{\mathbf{\Omega}^j}]\mathbf{y}. \qquad (A.57)$$

Since $\mathbf{x}$ is sparse with support $T$, $\mathbf{y}$ can be decomposed as

$$\mathbf{y} = \mathbf{\Phi}\mathbf{x} = \mathbf{\Phi}_{\mathbf{\Omega}^j}\mathbf{x}_{\mathbf{\Omega}^j} + \mathbf{\Phi}_{T\setminus\mathbf{\Omega}^j}\mathbf{x}_{T\setminus\mathbf{\Omega}^j}. \qquad (A.58)$$

From (A.57) and (A.58), it follows

---

3. An upper bound for $j_0 - 1$ can be found in [9].

$$\mathbf{r}^j = (\mathbf{I} - \mathbf{P}_{\mathbf{\Omega}^j})\mathbf{\Phi x} = (\mathbf{I} - \mathbf{P}_{\mathbf{\Omega}^j})\Big(\mathbf{\Phi}_{\mathbf{\Omega}^j}\mathbf{x}_{\mathbf{\Omega}^j} + \mathbf{\Phi}_{T\setminus\mathbf{\Omega}^j}\mathbf{x}_{T\setminus\mathbf{\Omega}^j}\Big) \stackrel{(a)}{=} (\mathbf{I} - \mathbf{P}_{\mathbf{\Omega}^j})\mathbf{\Phi}_{T\setminus\mathbf{\Omega}^j}\mathbf{x}_{T\setminus\mathbf{\Omega}^j} = (\mathbf{I} - \mathbf{P}_{\mathbf{\Omega}^j})\mathbf{\Phi}\tilde{\mathbf{x}}_{T\setminus\mathbf{\Omega}^j},$$
(A.59)

where (a) holds since $\mathbf{\Phi}_{\mathbf{\Omega}^j}\mathbf{x}_{\mathbf{\Omega}^j} \in \mathcal{R}(\mathbf{\Phi}_{\mathbf{\Omega}^j})$ and $\mathbf{I} - \mathbf{P}_{\mathbf{\Omega}^j}$ is the orthogonal projection onto the orthogonal complement of $\mathcal{R}(\mathbf{\Phi}_{\mathbf{\Omega}^j})$. With (A.59), we have

$$\|\mathbf{r}^j\|_2 = \|(\mathbf{I} - \mathbf{P}_{\mathbf{\Omega}^j})\mathbf{\Phi}\tilde{\mathbf{x}}_{T\setminus\mathbf{\Omega}^j}\|_2 \leq \|\mathbf{\Phi}\tilde{\mathbf{x}}_{T\setminus\mathbf{\Omega}^j}\|_2$$
$$\stackrel{(a)}{\leq} \sqrt{1+\delta_{3K}}\|\tilde{\mathbf{x}}_{T\setminus\mathbf{\Omega}^j}\|_2 = \sqrt{1+\delta_{3K}}\|\mathbf{x}_{T\setminus\mathbf{\Omega}^j}\|_2 \stackrel{(b)}{\leq} \sqrt{1+\delta_{3K}}\alpha\|\mathbf{x}_{T\setminus\mathbf{\Omega}^{j-1}}\|_2,$$
(A.60)

where (a) follows due to RIP and (b) holds from (A.56). Now, based on Step 3.8, $\mathbf{r}^j$ can also be written as

$$\mathbf{r}^j = \mathbf{y} - \mathbf{\Phi}_{\mathbf{\Omega}^j}\mathbf{q}_{\mathbf{\Omega}^j} = \underbrace{\mathbf{\Phi}_{T\setminus\mathbf{\Omega}^j}\mathbf{x}_{T\setminus\mathbf{\Omega}^j} + \mathbf{\Phi}_{\mathbf{\Omega}^j}\mathbf{x}_{\mathbf{\Omega}^j}}_{=\mathbf{\Phi x}} - \mathbf{\Phi}_{\mathbf{\Omega}^j}\mathbf{q}_{\mathbf{\Omega}^j} = \mathbf{\Phi}(\tilde{\mathbf{x}}_{T\setminus\mathbf{\Omega}^j} + \tilde{\mathbf{x}}_{\mathbf{\Omega}^j} - \tilde{\mathbf{q}}_{\mathbf{\Omega}^j}).$$
(A.61)

Then, we have

$$\|\mathbf{r}^{j-1}\|_2 = \|\mathbf{\Phi}(\tilde{\mathbf{x}}_{T\setminus\mathbf{\Omega}^{j-1}} + \tilde{\mathbf{x}}_{\mathbf{\Omega}^{j-1}} - \tilde{\mathbf{q}}_{\mathbf{\Omega}^{j-1}})\|_2$$
$$\stackrel{(a)}{\geq} \sqrt{1-\delta_{3K}}\|\tilde{\mathbf{x}}_{T\setminus\mathbf{\Omega}^{j-1}} + \tilde{\mathbf{x}}_{\mathbf{\Omega}^{j-1}} - \tilde{\mathbf{q}}_{\mathbf{\Omega}^{j-1}}\|_2$$
$$\stackrel{(b)}{=} \sqrt{1-\delta_{3K}}\sqrt{\|\tilde{\mathbf{x}}_{T\setminus\mathbf{\Omega}^{j-1}}\|_2^2 + \|\tilde{\mathbf{x}}_{\mathbf{\Omega}^{j-1}} - \tilde{\mathbf{q}}_{\mathbf{\Omega}^{j-1}}\|_2^2} \geq \sqrt{1-\delta_{3K}}\|\tilde{\mathbf{x}}_{T\setminus\mathbf{\Omega}^{j-1}}\|_2 = \sqrt{1-\delta_{3K}}\|\mathbf{x}_{T\setminus\mathbf{\Omega}^{j-1}}\|_2, \text{(A.62)}$$

where (a) follows from the RIP, and (b) is true since the supports of $\tilde{\mathbf{x}}_{T\setminus\mathbf{\Omega}^{j-1}}$ and $\tilde{\mathbf{x}}_{\mathbf{\Omega}^{j-1}} - \tilde{\mathbf{q}}_{\mathbf{\Omega}^{j-1}}$ are disjoint. One can verify that if $\delta_{3K} \leq 0.2412$, the following inequality holds:

$$\sqrt{1-\delta_{3K}} > \sqrt{1+\delta_{3K}}\alpha.$$
(A.63)

The assertion (A.54) then follows by combining (A.60), (A.62), and (A.63).

Now we turn to show that if $\|\mathbf{r}^{j_0-1}\|_2 = 0$ for some $j_0 - 1$, we also have $\|\mathbf{r}^{j_0}\|_2 = 0$. Since $\|\mathbf{r}^{j_0-1}\|_2 = 0$, thus $\mathbf{r}^{j_0-1} = \mathbf{0}$, it follows

$$\mathbf{0} = \mathbf{r}^{j_0-1} \stackrel{(a)}{=} \mathbf{y} - \mathbf{\Phi}_{\mathbf{\Omega}^{j_0-1}}\mathbf{q}_{\mathbf{\Omega}^{j_0-1}} = \mathbf{\Phi x} - \mathbf{\Phi}_{\mathbf{\Omega}^{j_0-1}}\mathbf{q}_{\mathbf{\Omega}^{j_0-1}} = \mathbf{\Phi x} - \mathbf{\Phi}\tilde{\mathbf{q}}_{\mathbf{\Omega}^{j_0-1}} = \mathbf{\Phi}(\mathbf{x} - \tilde{\mathbf{q}}_{\mathbf{\Omega}^{j_0-1}}).$$
(A.64)

where (a) holds according to Step 3.8 in Table II. Since $\mathbf{\Phi}$ satisfies RIP of order $3K$ and $\mathbf{x} - \tilde{\mathbf{q}}_{\mathbf{\Omega}^{j_0-1}}$ is $2K$-sparse, (A.64) immediately implies

$$\mathbf{x} = \tilde{\mathbf{q}}_{\mathbf{\Omega}^{j_0-1}}.$$
(A.65)

Since the supports of $\mathbf{x}$ and $\tilde{\mathbf{q}}_{\mathbf{\Omega}^{j_0-1}}$ are, respectively, $T$ and $\mathbf{\Omega}^{j_0-1}$, (A.65) then implies $\mathbf{\Omega}^{j_0-1} = T$, thereby

$$\|\mathbf{x}_{T\setminus\mathbf{\Omega}^{j_0-1}}\|_2 = 0.$$
(A.66)

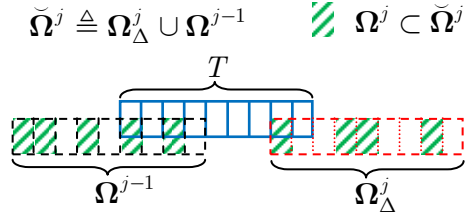

Figure 4. Depiction of the set $\breve{\boldsymbol{\Omega}}^j$.

Finally, it follows from (A.56) and (A.66) that $\left\|\mathbf{x}_{T\setminus\boldsymbol{\Omega}^{j_0}}\right\|_2 = 0$, which means the estimated support at the $j_0$-th iteration is still $\boldsymbol{\Omega}^{j_0} = T$. Then, $\mathbf{q}$ obtained by Steps 3.6-3.7 is exactly $\mathbf{x}$, and thus, $\mathbf{r}^{j_0} = \mathbf{0}$ (see Step 3.8). □

*[Proof of Lemma A.6]*:

First recall from the SP algorithm (see Table II) that $\boldsymbol{\Omega}_\Delta^j$ is the index set, with $|\boldsymbol{\Omega}_\Delta^j| = K$, such that $\left\{\left|(\boldsymbol{\Phi}^*\mathbf{r}^{j-1})_i\right|\right\}_{i\in\boldsymbol{\Omega}_\Delta^j}$ consists of the $K$ largest elements of $\left\{\left|(\boldsymbol{\Phi}^*\mathbf{r}^{j-1})_1\right|, \cdots, \left|(\boldsymbol{\Phi}^*\mathbf{r}^{j-1})_N\right|\right\}$, where $\mathbf{r}^{j-1}$ is the residual vector in the $(j-1)$-th iteration. To prove (A.56), it suffices to prove the following two conditions: For $\breve{\boldsymbol{\Omega}}^j = \boldsymbol{\Omega}^{j-1} \cup \boldsymbol{\Omega}_\Delta^j$ (see Figure 4 for a depiction of $\breve{\boldsymbol{\Omega}}^j$),

$$\left\|\mathbf{x}_{T\setminus\breve{\boldsymbol{\Omega}}^j}\right\|_2 \leq \frac{2\delta_{3K}}{1-\delta_{3K}}\sqrt{1+\delta_{3K}^2\frac{1+\delta_{3K}}{1-\delta_{3K}}}\left\|\mathbf{x}_{T\setminus\boldsymbol{\Omega}^{j-1}}\right\|_2 \qquad (A.67)$$

and

$$\left\|\mathbf{x}_{T\setminus\boldsymbol{\Omega}^j}\right\|_2 \leq \sqrt{1+4\delta_{3K}^2\frac{1+\delta_{3K}}{1-\delta_{3K}}}\left\|\mathbf{x}_{T\setminus\breve{\boldsymbol{\Omega}}^j}\right\|_2. \qquad (A.68)$$

Inequality (A.56) can be directly obtained from (A.67), (A.68), together with some straightforward manipulations.

*(i) Derivations of (A.67):*

We first observe that, in the $j$-th iteration, $\mathbf{r}^{j-1}$ needed in Step 3.1 can be expressed as

$$\mathbf{r}^{j-1} \stackrel{(a)}{=} \mathbf{y} - \boldsymbol{\Phi}_{\boldsymbol{\Omega}^{j-1}}\mathbf{q}_{\boldsymbol{\Omega}^{j-1}} = \mathbf{y} - \mathbf{P}_{\boldsymbol{\Omega}^{j-1}}\mathbf{y} = (\mathbf{I} - \mathbf{P}_{\boldsymbol{\Omega}^{j-1}})\mathbf{y} = (\mathbf{I} - \mathbf{P}_{\boldsymbol{\Omega}^{j-1}})\boldsymbol{\Phi}_T\mathbf{x}_T$$
$$= (\mathbf{I} - \mathbf{P}_{\boldsymbol{\Omega}^{j-1}})(\boldsymbol{\Phi}_{T\setminus\boldsymbol{\Omega}^{j-1}}\mathbf{x}_{T\setminus\boldsymbol{\Omega}^{j-1}} + \boldsymbol{\Phi}_{\boldsymbol{\Omega}^{j-1}}\mathbf{x}_{\boldsymbol{\Omega}^{j-1}}) \stackrel{(b)}{=} (\mathbf{I} - \mathbf{P}_{\boldsymbol{\Omega}^{j-1}})\boldsymbol{\Phi}_{T\setminus\boldsymbol{\Omega}^{j-1}}\mathbf{x}_{T\setminus\boldsymbol{\Omega}^{j-1}}, \qquad (A.69)$$

where (a) is due to Step 3.8 of the SP algorithm and (b) follows since $\boldsymbol{\Phi}_{\boldsymbol{\Omega}^{j-1}}\mathbf{x}_{\boldsymbol{\Omega}^{j-1}} \in \mathcal{R}(\boldsymbol{\Phi}_{\boldsymbol{\Omega}^{j-1}})$ and $\mathbf{I} - \mathbf{P}_{\boldsymbol{\Omega}^{j-1}}$ is the orthogonal projection onto the orthogonal complement of $\mathcal{R}(\boldsymbol{\Phi}_{\boldsymbol{\Omega}^{j-1}})$. Also, we note that there exists one $\mathbf{z} \in \mathbb{R}^{|\boldsymbol{\Omega}^{j-1}|}$ such that

$$\mathbf{P}_{\boldsymbol{\Omega}^{j-1}}\boldsymbol{\Phi}_{T\setminus\boldsymbol{\Omega}^{j-1}}\mathbf{x}_{T\setminus\boldsymbol{\Omega}^{j-1}} = \boldsymbol{\Phi}_{\boldsymbol{\Omega}^{j-1}}\mathbf{z}, \qquad (A.70)$$

and thus,

$$\left\|\mathbf{P}_{\mathbf{\Omega}^{j-1}}\mathbf{\Phi}_{T\setminus\mathbf{\Omega}^{j-1}}\mathbf{x}_{T\setminus\mathbf{\Omega}^{j-1}}\right\|_2 = \left\|\mathbf{\Phi}_{\mathbf{\Omega}^{j-1}}\mathbf{z}\right\|_2 \overset{(a)}{\geq} \sqrt{1-\delta_{3K}}\,\|\mathbf{z}\|_2, \qquad (A.71)$$

where (a) is due to Lemma A.1. Also, based on Lemma A.4, we have

$$\left\|\mathbf{P}_{\mathbf{\Omega}^{j-1}}\mathbf{\Phi}_{T\setminus\mathbf{\Omega}^{j-1}}\mathbf{x}_{T\setminus\mathbf{\Omega}^{j-1}}\right\|_2 = \left\|\mathbf{P}_{\mathbf{\Omega}^{j-1}}\mathbf{\Phi}\tilde{\mathbf{x}}_{T\setminus\mathbf{\Omega}^{j-1}}\right\|_2 \leq \delta_{3K}\sqrt{1+\delta_{3K}}\left\|\tilde{\mathbf{x}}_{T\setminus\mathbf{\Omega}^{j-1}}\right\|_2 \leq \delta_{3K}\sqrt{1+\delta_{3K}}\left\|\mathbf{x}_{T\setminus\mathbf{\Omega}^{j-1}}\right\|_2. (A.72)$$

It follows directly from (A.71) and (A.72) that

$$\|\mathbf{z}\|_2 \leq \delta_{3K}\frac{\sqrt{1+\delta_{3K}}}{\sqrt{1-\delta_{3K}}}\left\|\mathbf{x}_{T\setminus\mathbf{\Omega}^{j-1}}\right\|_2. \qquad (A.73)$$

By means of (A.69) and (A.70), we can further rewrite the residual $\mathbf{r}^{j-1}$ as

$$\mathbf{r}^{j-1} = \mathbf{\Phi}_{T\setminus\mathbf{\Omega}^{j-1}}\mathbf{x}_{T\setminus\mathbf{\Omega}^{j-1}} - \mathbf{\Phi}_{\mathbf{\Omega}^{j-1}}\mathbf{z} = \mathbf{\Phi}\tilde{\mathbf{x}}_{T\setminus\mathbf{\Omega}^{j-1}} - \mathbf{\Phi}\tilde{\mathbf{z}} = \mathbf{\Phi}\underbrace{(\tilde{\mathbf{x}}_{T\setminus\mathbf{\Omega}^{j-1}} - \tilde{\mathbf{z}})}_{\triangleq \mathbf{h}^{j-1}} = \mathbf{\Phi}_{T\cup\mathbf{\Omega}^{j-1}}\mathbf{h}^{j-1}_{T\cup\mathbf{\Omega}^{j-1}}, \quad (A.74)$$

where $\tilde{\mathbf{z}}$ is obtained by padding zeros to $\mathbf{z}$ (the support of $\tilde{\mathbf{z}}$ is $\mathbf{\Omega}^{j-1}$). According to Step 3.1 of the SP algorithm (see Table II), we must have

$$\left\|\mathbf{\Phi}_S^*\mathbf{r}^{j-1}\right\|_2 \leq \left\|\mathbf{\Phi}_{\mathbf{\Omega}_\Delta^j}^*\mathbf{r}^{j-1}\right\|_2 \qquad (A.75)$$

for any subset $S \subset \{1,\ldots,N\}$ of $K$ elements, and thus

$$\left\|\mathbf{\Phi}_T^*\mathbf{r}^{j-1}\right\|_2^2 \leq \left\|\mathbf{\Phi}_{\mathbf{\Omega}_\Delta^j}^*\mathbf{r}^{j-1}\right\|_2^2. \qquad (A.76)$$

Since $\left\|\mathbf{\Phi}_T^*\mathbf{r}^{j-1}\right\|_2^2 = \left\|\mathbf{\Phi}_{T\setminus\mathbf{\Omega}_\Delta^j}^*\mathbf{r}^{j-1}\right\|_2^2 + \left\|\mathbf{\Phi}_{T\cap\mathbf{\Omega}_\Delta^j}^*\mathbf{r}^{j-1}\right\|_2^2$ and $\left\|\mathbf{\Phi}_{\mathbf{\Omega}_\Delta^j}^*\mathbf{r}^{j-1}\right\|_2^2 = \left\|\mathbf{\Phi}_{\mathbf{\Omega}_\Delta^j\setminus T}^*\mathbf{r}^{j-1}\right\|_2^2 + \left\|\mathbf{\Phi}_{\mathbf{\Omega}_\Delta^j\cap T}^*\mathbf{r}^{j-1}\right\|_2^2$, inequality (A.76) then implies

$$\left\|\mathbf{\Phi}_{T\setminus\mathbf{\Omega}_\Delta^j}^*\mathbf{r}^{j-1}\right\|_2^2 \leq \left\|\mathbf{\Phi}_{\mathbf{\Omega}_\Delta^j\setminus T}^*\mathbf{r}^{j-1}\right\|_2^2. \qquad (A.77)$$

An upper bound of the right-hand side (RHS) of (A.77) is derived as

$$\left\|\mathbf{\Phi}_{\mathbf{\Omega}_\Delta^j\setminus T}^*\mathbf{r}^{j-1}\right\|_2 \overset{(a)}{=} \left\|\mathbf{\Phi}_{\mathbf{\Omega}_\Delta^j\setminus T}^*\mathbf{\Phi}_{T\cup\mathbf{\Omega}^{j-1}}\mathbf{h}^{j-1}_{T\cup\mathbf{\Omega}^{j-1}}\right\|_2 \overset{(b)}{\leq} \delta_{3K}\left\|\mathbf{h}^{j-1}_{T\cup\mathbf{\Omega}^{j-1}}\right\|_2 \leq \delta_{3K}\left\|\mathbf{h}^{j-1}\right\|_2, \qquad (A.78)$$

where (a) follows from (A.74), and (b) holds since $\mathbf{\Omega}_\Delta^j \cap \mathbf{\Omega}^{j-1} = \phi$, the empty set, and using Lemma A.5. Also, a lower bound of the left-hand side (LHS) of (A.77) is obtained as

$$\left\|\mathbf{\Phi}_{T\setminus\mathbf{\Omega}_{\Delta}^{j}}^{*}\mathbf{r}^{j-1}\right\|_{2} \overset{(a)}{\geq} \left\|\mathbf{\Phi}_{T\setminus\breve{\mathbf{\Omega}}^{j}}^{*}\mathbf{r}^{j-1}\right\|_{2}$$

$$\overset{(b)}{=} \left\|\mathbf{\Phi}_{T\setminus\breve{\mathbf{\Omega}}^{j}}^{*}\mathbf{\Phi}\mathbf{h}^{j-1}\right\|_{2}$$

$$\overset{(c)}{=} \left\|\mathbf{\Phi}_{T\setminus\breve{\mathbf{\Omega}}^{j}}^{*}\left[\mathbf{\Phi}_{T\setminus\breve{\mathbf{\Omega}}^{j}}\mathbf{h}_{T\setminus\breve{\mathbf{\Omega}}^{j}}^{j-1} + \mathbf{\Phi}_{\breve{\mathbf{\Omega}}^{j}}\mathbf{h}_{\breve{\mathbf{\Omega}}^{j}}^{j-1}\right]\right\|_{2} \geq \left\|\mathbf{\Phi}_{T\setminus\breve{\mathbf{\Omega}}^{j}}^{*}\mathbf{\Phi}_{T\setminus\breve{\mathbf{\Omega}}^{j}}\mathbf{h}_{T\setminus\breve{\mathbf{\Omega}}^{j}}^{j-1}\right\|_{2} - \left\|\mathbf{\Phi}_{T\setminus\breve{\mathbf{\Omega}}^{j}}^{*}\mathbf{\Phi}_{\breve{\mathbf{\Omega}}^{j}}\mathbf{h}_{\breve{\mathbf{\Omega}}^{j}}^{j-1}\right\|_{2}$$

$$\overset{(d)}{\geq} \left\|\mathbf{\Phi}_{T\setminus\breve{\mathbf{\Omega}}^{j}}^{*}\mathbf{\Phi}_{T\setminus\breve{\mathbf{\Omega}}^{j}}\mathbf{h}_{T\setminus\breve{\mathbf{\Omega}}^{j}}^{j-1}\right\|_{2} - \delta_{3K}\left\|\mathbf{h}_{\breve{\mathbf{\Omega}}^{j}}^{j-1}\right\|_{2}$$

$$\overset{(e)}{\geq} (1-\delta_{3K})\left\|\mathbf{h}_{T\setminus\breve{\mathbf{\Omega}}^{j}}^{j-1}\right\|_{2} - \delta_{3K}\left\|\mathbf{h}_{\breve{\mathbf{\Omega}}^{j}}^{j-1}\right\|_{2}$$

$$\overset{(f)}{\geq} (1-\delta_{3K})\left\|\mathbf{x}_{T\setminus\breve{\mathbf{\Omega}}^{j}}\right\|_{2} - \delta_{3K}\left\|\mathbf{h}^{j-1}\right\|_{2}, \tag{A.79}$$

where (a) follows because $\mathbf{\Omega}_{\Delta}^{j} \subset \breve{\mathbf{\Omega}}^{j}(=\mathbf{\Omega}^{j-1}\cup\mathbf{\Omega}_{\Delta}^{j})$, (b) is obtained using (A.74), (c) holds since the support of $\mathbf{h}^{j-1} = \tilde{\mathbf{x}}_{T\setminus\mathbf{\Omega}^{j-1}} - \tilde{\mathbf{z}}$ is $T\cup\mathbf{\Omega}^{j-1}$ and $(T\cup\mathbf{\Omega}^{j-1}) \subset (T\cup\breve{\mathbf{\Omega}}^{j})$, (d) is obtained by using Lemmas A.5, (e) holds by means of Lemma A.3, and (f) holds since $\mathbf{h}_{T\setminus\breve{\mathbf{\Omega}}^{j}}^{j-1} = (\tilde{\mathbf{x}}_{T\setminus\mathbf{\Omega}^{j-1}} - \tilde{\mathbf{z}})_{T\setminus\breve{\mathbf{\Omega}}^{j}} = \mathbf{x}_{T\setminus\breve{\mathbf{\Omega}}^{j}}$ and $\left\|\mathbf{h}_{\breve{\mathbf{\Omega}}^{j}}^{j-1}\right\|_{2} \leq \left\|\mathbf{h}^{j-1}\right\|_{2}$. Now, we can reach the following

$$\left\|\mathbf{x}_{T\setminus\breve{\mathbf{\Omega}}^{j}}\right\|_{2} \overset{(a)}{\leq} \frac{2\delta_{3K}}{1-\delta_{3K}}\left\|\mathbf{h}^{j-1}\right\|_{2}$$

$$\overset{(b)}{=} \frac{2\delta_{3K}}{1-\delta_{3K}}\sqrt{\left\|\tilde{\mathbf{x}}_{T\setminus\mathbf{\Omega}^{j-1}}\right\|_{2}^{2} + \|\tilde{\mathbf{z}}\|_{2}^{2}} = \frac{2\delta_{3K}}{1-\delta_{3K}}\sqrt{\left\|\mathbf{x}_{T\setminus\mathbf{\Omega}^{j-1}}\right\|_{2}^{2} + \|\mathbf{z}\|_{2}^{2}} \tag{A.80}$$

$$\overset{(c)}{\leq} \frac{2\delta_{3K}}{1-\delta_{3K}}\sqrt{1+\delta_{3K}^{2}\frac{1+\delta_{3K}}{1-\delta_{3K}}}\left\|\mathbf{x}_{T\setminus\mathbf{\Omega}^{j-1}}\right\|_{2},$$

where (a) holds by combining (A.77), (A.78) and (A.79), (b) is true since $\mathbf{h}^{j-1} = \tilde{\mathbf{x}}_{T\setminus\mathbf{\Omega}^{j-1}} - \tilde{\mathbf{z}}$ (see (A.74)) and the supports of $\tilde{\mathbf{x}}_{T\setminus\mathbf{\Omega}^{j}}$ and $\tilde{\mathbf{z}}$ are disjoint, and (c) follows from (A.73). Hence, (A.67) is proved.

*(ii) Derivations of (A.68):*

According to the SP algorithm (see Step 3.5 in Table II), the $K$ elements of the estimated support $\mathbf{\Omega}^{j} \subset \breve{\mathbf{\Omega}}^{j}$ are the $K$ indexes corresponding to the $K$ largest magnitudes of $\breve{\mathbf{q}}$. Since the entries of $\breve{\mathbf{q}}$ are all zeros except those $2K$ elements indexed by $\breve{\mathbf{\Omega}}^{j}$ (see Steps 3.3-3.4 in Table II), the entries of $\breve{\mathbf{q}}$ indexed by $\bar{\mathbf{\Omega}}^{j} \triangleq \breve{\mathbf{\Omega}}^{j} \setminus \mathbf{\Omega}^{j}$ are thus those yielding the $K$ smallest nonzero magnitudes; more precisely, we have

$$\left\|\breve{\mathbf{q}}_{\bar{\mathbf{\Omega}}^{j}}\right\|_{2} \leq \left\|\breve{\mathbf{q}}_{S_{2}}\right\|_{2} \text{ for all } S_{2} \subset \breve{\mathbf{\Omega}}^{j} \text{ of } K \text{ (or more) elements.} \tag{A.81}$$

Since $(\breve{\mathbf{\Omega}}^{j} \setminus T) \subset \breve{\mathbf{\Omega}}^{j}$ consists of $K$ or more elements, it can be deduced from (A.81) that

$$\left\|\breve{\mathbf{q}}_{\bar{\Omega}^j}\right\|_2 \le \left\|\breve{\mathbf{q}}_{\bar{\Omega}^j\setminus T}\right\|_2. \tag{A.82}$$

We claim that

$$\left\|\breve{\mathbf{q}}_{\bar{\Omega}^j\setminus T}\right\|_2 \le \delta_{3K}\frac{\sqrt{1+\delta_{3K}}}{\sqrt{1-\delta_{3K}}}\left\|\mathbf{x}_{T\setminus\bar{\Omega}^j}\right\|_2, \tag{A.83}$$

and

$$\left\|\breve{\mathbf{q}}_{\bar{\Omega}^j}\right\|_2 \ge \sqrt{\left\|\mathbf{x}_{T\setminus\Omega^j}\right\|_2^2 - \left\|\mathbf{x}_{T\setminus\bar{\Omega}^j}\right\|_2^2} - \delta_{3K}\frac{\sqrt{1+\delta_{3K}}}{\sqrt{1-\delta_{3K}}}\left\|\mathbf{x}_{T\setminus\bar{\Omega}^j}\right\|_2. \tag{A.84}$$

Then, inequality (A.68) can be obtained based on (A.82)~(A.84) together with some straightforward manipulations.

To prove (A.83), we note that, since $\mathbf{y} = \mathbf{\Phi}\mathbf{x}$ with $\mathbf{x}$ supported on $T$, it follows

$$\mathbf{P}_{\bar{\Omega}^j}\mathbf{y} = \mathbf{P}_{\bar{\Omega}^j}(\mathbf{\Phi}_{T\setminus\bar{\Omega}^j}\mathbf{x}_{T\setminus\bar{\Omega}^j} + \mathbf{\Phi}_{\bar{\Omega}^j}\mathbf{x}_{\bar{\Omega}^j}) = \mathbf{P}_{\bar{\Omega}^j}\mathbf{\Phi}_{T\setminus\bar{\Omega}^j}\mathbf{x}_{T\setminus\bar{\Omega}^j} + \mathbf{\Phi}_{\bar{\Omega}^j}\mathbf{x}_{\bar{\Omega}^j}. \tag{A.85}$$

Since there exists one $\mathbf{z} \in \mathbb{R}^{2K}$ such that

$$\mathbf{P}_{\bar{\Omega}^j}\mathbf{\Phi}_{T\setminus\bar{\Omega}^j}\mathbf{x}_{T\setminus\bar{\Omega}^j} = \mathbf{\Phi}_{\bar{\Omega}^j}\mathbf{z}, \tag{A.86}$$

(A.85) becomes

$$\mathbf{P}_{\bar{\Omega}^j}\mathbf{y} = \mathbf{\Phi}_{\bar{\Omega}^j}(\mathbf{z} + \mathbf{x}_{\bar{\Omega}^j}) = \mathbf{\Phi}(\tilde{\mathbf{z}} + \tilde{\mathbf{x}}_{\bar{\Omega}^j}) \stackrel{(a)}{=} \mathbf{\Phi}\underbrace{(\tilde{\mathbf{z}} + \tilde{\mathbf{x}}_{T\cap\bar{\Omega}^j})}_{\breve{\mathbf{q}}\text{ in Step 3.5}}; \tag{A.87}$$

in (A.87) $\tilde{\mathbf{z}}$ is obtained by padding zeros to $\mathbf{z}$ (the support of $\tilde{\mathbf{z}}$ is $\breve{\Omega}^j$), and (a) follows since only those entries of $\mathbf{x}$ indexed by $T$ are nonzero. Then, we obtain the following

$$\left\|\breve{\mathbf{q}}_{\bar{\Omega}^j\setminus T}\right\|_2 \stackrel{(a)}{=} \left\|(\tilde{\mathbf{z}} + \tilde{\mathbf{x}}_{T\cap\bar{\Omega}^j})_{\bar{\Omega}^j\setminus T}\right\|_2 = \left\|\tilde{\mathbf{z}}_{\bar{\Omega}^j\setminus T}\right\|_2 \le \|\tilde{\mathbf{z}}\|_2, \tag{A.88}$$

where (a) follows from (A.87). Also, we have

$$\sqrt{1-\delta_{3K}}\|\tilde{\mathbf{z}}\|_2 \stackrel{(a)}{\le} \|\mathbf{\Phi}\tilde{\mathbf{z}}\|_2 = \left\|\mathbf{\Phi}_{\bar{\Omega}^j}\mathbf{z}\right\|_2 \stackrel{(b)}{=} \left\|\mathbf{P}_{\bar{\Omega}^j}(\mathbf{\Phi}_{T\setminus\bar{\Omega}^j}\mathbf{x}_{T\setminus\bar{\Omega}^j})\right\|_2 \stackrel{(c)}{\le} \delta_{3K}\sqrt{1+\delta_{3K}}\left\|\mathbf{x}_{T\setminus\bar{\Omega}^j}\right\|_2, \tag{A.89}$$

where (a) follows since $\mathbf{\Phi}$ satisfies RIP, (b) follows from (A.86) and (c) is due to Lemma A.4. Inequality (A.83) directly follows from (A.88) and (A.89).

To prove (A.84), we first observe that

$$\left\|\breve{\mathbf{q}}_{\bar{\Omega}^j}\right\|_2 \stackrel{(a)}{=} \left\|(\tilde{\mathbf{z}} + \tilde{\mathbf{x}}_{T\cap\bar{\Omega}^j})_{\bar{\Omega}^j}\right\|_2 \ge \left\|(\tilde{\mathbf{x}}_{T\cap\bar{\Omega}^j})_{\bar{\Omega}^j}\right\|_2 - \left\|(\tilde{\mathbf{z}})_{\bar{\Omega}^j}\right\|_2$$
$$\ge \left\|(\tilde{\mathbf{x}}_{T\cap\bar{\Omega}^j})_{\bar{\Omega}^j}\right\|_2 - \|\tilde{\mathbf{z}}\|_2 \stackrel{(b)}{=} \left\|(\tilde{\mathbf{x}}_{\bar{\Omega}^j})_{\bar{\Omega}^j}\right\|_2 - \|\tilde{\mathbf{z}}\|_2 \stackrel{(c)}{=} \left\|\tilde{\mathbf{x}}_{\bar{\Omega}^j}\right\|_2 - \|\tilde{\mathbf{z}}\|_2, \tag{A.90}$$

where (a) follows from (A.87), (b) is true because $\mathbf{x}$ is supported on $T$, and (c) follows since $\bar{\Omega}^j = \breve{\Omega}^j \setminus \Omega^j \subset \breve{\Omega}^j$. Also, note that

$$\tilde{\mathbf{x}}_{T\setminus\Omega^j} = \tilde{\mathbf{x}}_{T\setminus(\breve{\Omega}^j\setminus\bar{\Omega}^j)} \stackrel{(a)}{=} \tilde{\mathbf{x}}_{(T\setminus\breve{\Omega}^j)\cup(T\cap\bar{\Omega}^j)} \stackrel{(b)}{=} \tilde{\mathbf{x}}_{T\setminus\breve{\Omega}^j} + \tilde{\mathbf{x}}_{T\cap\bar{\Omega}^j}, \tag{A.91}$$

where (a) holds since $T \setminus (\breve{\boldsymbol{\Omega}}^j \setminus \bar{\boldsymbol{\Omega}}^j) = T \setminus (\breve{\boldsymbol{\Omega}}^j \cap (\bar{\boldsymbol{\Omega}}^j)^c) = (T \setminus \breve{\boldsymbol{\Omega}}^j) \cup (T \cap \bar{\boldsymbol{\Omega}}^j)$ and (b) follows since $\bar{\boldsymbol{\Omega}}^j \subset \breve{\boldsymbol{\Omega}}^j$ and, therefore, $\tilde{\mathbf{x}}_{T \setminus \breve{\boldsymbol{\Omega}}^j}$ and $\tilde{\mathbf{x}}_{T \cap \bar{\boldsymbol{\Omega}}^j}$ have disjoint supports. By means of (A.91), we obtain

$$\left\| \tilde{\mathbf{x}}_{T \setminus \boldsymbol{\Omega}^j} \right\|_2^2 = \left\| \tilde{\mathbf{x}}_{T \setminus \bar{\boldsymbol{\Omega}}^j} \right\|_2^2 + \left\| \tilde{\mathbf{x}}_{T \cap \bar{\boldsymbol{\Omega}}^j} \right\|_2^2 \overset{(a)}{=} \left\| \tilde{\mathbf{x}}_{T \setminus \bar{\boldsymbol{\Omega}}^j} \right\|_2^2 + \left\| \tilde{\mathbf{x}}_{\bar{\boldsymbol{\Omega}}^j} \right\|_2^2 \tag{A.92}$$

where (a) follows because $\mathbf{x}$ is supported on $T$. With (A.92) we directly have

$$\left\| \tilde{\mathbf{x}}_{\bar{\boldsymbol{\Omega}}^j} \right\|_2 = \sqrt{\left\| \tilde{\mathbf{x}}_{T \setminus \boldsymbol{\Omega}^j} \right\|_2^2 - \left\| \tilde{\mathbf{x}}_{T \setminus \breve{\boldsymbol{\Omega}}^j} \right\|_2^2} = \sqrt{\left\| \mathbf{x}_{T \setminus \boldsymbol{\Omega}^j} \right\|_2^2 - \left\| \mathbf{x}_{T \setminus \breve{\boldsymbol{\Omega}}^j} \right\|_2^2}. \tag{A.93}$$

The assertion (A.84) thus holds by combining (A.89), (A.90) and (A.93). $\square$

*E. Proof of Theorem 4.8*

We need the next lemma to complete the proof. In what follows, assume that the SP algorithm terminates after $l$ iterations[4].

***Lemma A.7:*** Under the assumptions as in Theorem 4.8, it follows that

$$\left\| \mathbf{x}_{T \setminus \boldsymbol{\Omega}^l} \right\|_2 \leq \frac{(2 + \sqrt{1 + \delta_{3K}} \beta) \|\mathbf{w}\|_2}{\sqrt{1 - \delta_{3K}} - \sqrt{1 + \delta_{3K}} \alpha}, \tag{A.94}$$

where $\alpha$ and $\beta$ are defined in (4.17) and (4.18).

*[Proof]:* The proof is relegated to the end of this appendix. $\square$

Since $\hat{\mathbf{x}} = \mathbf{q}$ is supported by $\boldsymbol{\Omega}^l$ (see Step 4 of SP), the reconstruction error is

$$\|\mathbf{x} - \hat{\mathbf{x}}\|_2^2 = \left\| \underbrace{\tilde{\mathbf{x}}_{T \setminus \boldsymbol{\Omega}^l} + \tilde{\mathbf{x}}_{\boldsymbol{\Omega}^l}}_{=\mathbf{x}} - \hat{\mathbf{x}} \right\|_2^2 \overset{(a)}{=} \left\| \tilde{\mathbf{x}}_{T \setminus \boldsymbol{\Omega}^l} \right\|_2^2 + \left\| \tilde{\mathbf{x}}_{\boldsymbol{\Omega}^l} - \hat{\mathbf{x}} \right\|_2^2 = \left\| \tilde{\mathbf{x}}_{T \setminus \boldsymbol{\Omega}^l} \right\|_2^2 + \left\| \tilde{\mathbf{x}}_{\boldsymbol{\Omega}^l} - \mathbf{q} \right\|_2^2$$

$$= \left\| \mathbf{x}_{T \setminus \boldsymbol{\Omega}^l} \right\|_2^2 + \left\| \mathbf{x}_{\boldsymbol{\Omega}^l} - \mathbf{q}_{\boldsymbol{\Omega}^l} \right\|_2^2, \tag{A.95}$$

where (a) is true since the supports of $\tilde{\mathbf{x}}_{T \setminus \boldsymbol{\Omega}^l}$ and $\tilde{\mathbf{x}}_{\boldsymbol{\Omega}^l} - \hat{\mathbf{x}}$ are, respectively, $T \setminus \boldsymbol{\Omega}^l$ and $\boldsymbol{\Omega}^l$, which are disjoint. By the definition of $\mathbf{q}_{\boldsymbol{\Omega}^l}$ (see Step 3.7 of SP), we have

$$\begin{aligned}\boldsymbol{\Phi}_{\Omega^l}\mathbf{q}_{\Omega^l} &= \mathbf{P}_{\Omega^l}\mathbf{y} = \mathbf{P}_{\Omega^l}\boldsymbol{\Phi}\mathbf{x} + \mathbf{P}_{\Omega^l}\mathbf{w} = \mathbf{P}_{\Omega^l}\boldsymbol{\Phi}_{\Omega^l}\mathbf{x}_{\Omega^l} + \mathbf{P}_{\Omega^l}\boldsymbol{\Phi}_{T\setminus\Omega^l}\mathbf{x}_{T\setminus\Omega^l} + \mathbf{P}_{\Omega^l}\mathbf{w}\\ &= \boldsymbol{\Phi}_{\Omega^l}\mathbf{x}_{\Omega^l} + \underbrace{\mathbf{P}_{\Omega^l}\boldsymbol{\Phi}_{T\setminus\Omega^l}\mathbf{x}_{T\setminus\Omega^l}}_{\triangleq \boldsymbol{\Phi}_{\Omega^l}\mathbf{f}} + \underbrace{\mathbf{P}_{\Omega^l}\mathbf{w}}_{\triangleq \boldsymbol{\Phi}_{\Omega^l}\mathbf{g}} = \boldsymbol{\Phi}_{\Omega^l}(\mathbf{x}_{\Omega^l} + \mathbf{f} + \mathbf{g}).\end{aligned} \quad (A.96)$$

---

4. Since there are totally $C_K^N = \dfrac{N!}{K!(N-K)!}$ candidate supports, the SP algorithm with the stopping criterion $\|\mathbf{r}^j\|_2 \geq \|\mathbf{r}^{j-1}\|_2$ halts after at most $C_K^N + 1$ iterations. This is because $\boldsymbol{\Omega}^{C_K^N + 1}$, the support identified in the $(C_K^N + 1)$-th iteration, must be the same as $\boldsymbol{\Omega}^{j_0}$, the support identified in the $j_0$-th iteration for some $1 \leq j_0 \leq C_K^N$. As a result, in the worst case we have $\left\| \mathbf{r}^{C_K^N + 1} \right\|_2 = \left\| \mathbf{r}^{j_0} \right\|_2 > \left\| \mathbf{r}^{j_0 + 1} \right\|_2 > \cdots > \left\| \mathbf{r}^{C_K^N} \right\|_2$.

Since $\boldsymbol{\Phi}$ satisfies RIP of order $3K$, arbitrary $3K$ columns of $\boldsymbol{\Phi}$ are linearly independent, and so are those of $\boldsymbol{\Phi}_{\Omega^l}$. Thus, we have from (A.96) that

$$\mathbf{q}_{\Omega^l} = \mathbf{x}_{\Omega^l} + \mathbf{f} + \mathbf{g}. \quad (A.97)$$

Based on (A.95) and (A.97), we have

$$\|\mathbf{x} - \hat{\mathbf{x}}\|_2^2 = \left\|\mathbf{x}_{T\setminus\Omega^l}\right\|_2^2 + \|\mathbf{f} + \mathbf{g}\|_2^2 \leq \left\|\mathbf{x}_{T\setminus\Omega^l}\right\|_2^2 + (\|\mathbf{f}\|_2 + \|\mathbf{g}\|_2)^2. \quad (A.98)$$

Note that

$$\sqrt{1-\delta_{3K}}\|\mathbf{f}\|_2 \overset{(a)}{\leq} \left\|\boldsymbol{\Phi}_{\Omega^l}\mathbf{f}\right\|_2 \overset{(b)}{=} \left\|\mathbf{P}_{\Omega^l}\boldsymbol{\Phi}_{T\setminus\Omega^l}\mathbf{x}_{T\setminus\Omega^l}\right\|_2 = \left\|\mathbf{P}_{\Omega^l}\boldsymbol{\Phi}\tilde{\mathbf{x}}_{T\setminus\Omega^l}\right\|_2 \overset{(c)}{\leq} \delta_{3K}\sqrt{1+\delta_{3K}}\left\|\tilde{\mathbf{x}}_{T\setminus\Omega^l}\right\|_2$$
$$= \delta_{3K}\sqrt{1+\delta_{3K}}\left\|\mathbf{x}_{T\setminus\Omega^l}\right\|_2, \quad (A.99)$$

where (a) follows from Lemma A.1, (b) holds by the definition of $\boldsymbol{\Phi}_{\Omega^l}\mathbf{f}$ in (A.96), and (c) can be obtained using Lemma A.4. By the definition of $\mathbf{g}$ (see (A.96)), it follows with the aid of Lemma A.1 that

$$\sqrt{1-\delta_{3K}}\|\mathbf{g}\|_2 \leq \left\|\boldsymbol{\Phi}_{\Omega^l}\mathbf{g}\right\|_2 = \left\|\mathbf{P}_{\Omega^l}\mathbf{w}\right\|_2 \leq \|\mathbf{w}\|_2. \quad (A.100)$$

Based on (A.98)~(A.100), it follows that

$$\begin{aligned}\|\mathbf{x} - \hat{\mathbf{x}}\|_2^2 &\leq \left\|\mathbf{x}_{T\setminus\Omega^l}\right\|_2^2 + \left(\frac{\delta_{3K}\sqrt{1+\delta_{3K}}}{\sqrt{1-\delta_{3K}}}\left\|\mathbf{x}_{T\setminus\Omega^l}\right\|_2 + \frac{1}{\sqrt{1-\delta_{3K}}}\|\mathbf{w}\|_2\right)^2\\ &\leq \left((1+\frac{\delta_{3K}\sqrt{1+\delta_{3K}}}{\sqrt{1-\delta_{3K}}})\left\|\mathbf{x}_{T\setminus\Omega^l}\right\|_2 + \frac{1}{\sqrt{1-\delta_{3K}}}\|\mathbf{w}\|_2\right)^2.\end{aligned} \quad (A.101)$$

Using Lemma A.7 and (A.101) together with some straightforward manipulations, the assertion in Theorem 4.8 thus follows. $\square$

*[Proof of Lemma A.7]*:

According to Step 3.8 of SP (see Table II), we have

$$\begin{aligned}\mathbf{r}^j &= \mathbf{y} - \boldsymbol{\Phi}_{\Omega^j}\mathbf{q}_{\Omega^j} = (\mathbf{I} - \mathbf{P}_{\Omega^j})\mathbf{y} = (\mathbf{I} - \mathbf{P}_{\Omega^j})(\boldsymbol{\Phi}\mathbf{x} + \mathbf{w})\\ \mathbf{r}^j &= (\mathbf{I} - \mathbf{P}_{\Omega^j})(\boldsymbol{\Phi}_{\Omega^j}\mathbf{x}_{\Omega^j} + \boldsymbol{\Phi}_{T\setminus\Omega^j}\mathbf{x}_{T\setminus\Omega^j} + \mathbf{w}) = (\mathbf{I} - \mathbf{P}_{\Omega^j})(\boldsymbol{\Phi}_{T\setminus\Omega^j}\mathbf{x}_{T\setminus\Omega^j} + \mathbf{w}).\end{aligned} \quad (A.102)$$

Based on (A.102), it follows

$$\left\|\mathbf{r}^{j}\right\|_{2} = \left\|(\mathbf{I}-\mathbf{P}_{\mathbf{\Omega}^{j}})(\mathbf{\Phi}_{T\backslash\mathbf{\Omega}^{j}}\mathbf{x}_{T\backslash\mathbf{\Omega}^{j}} + \mathbf{w})\right\|_{2} \leq \left\|\mathbf{\Phi}_{T\backslash\mathbf{\Omega}^{j}}\mathbf{x}_{T\backslash\mathbf{\Omega}^{j}} + \mathbf{w}\right\|_{2} \leq \left\|\mathbf{\Phi}_{T\backslash\mathbf{\Omega}^{j}}\mathbf{x}_{T\backslash\mathbf{\Omega}^{j}}\right\|_{2} + \left\|\mathbf{w}\right\|_{2}$$
$$\stackrel{(a)}{\leq} \sqrt{1+\delta_{3K}}\left\|\mathbf{x}_{T\backslash\mathbf{\Omega}^{j}}\right\|_{2} + \left\|\mathbf{w}\right\|_{2}, \tag{A.103}$$

where (a) is due to Lemma A.1. Again from Step 3.8, $\mathbf{r}^{j}$ can also be written as $\mathbf{r}^{j} = \mathbf{y} - \mathbf{\Phi}_{\mathbf{\Omega}^{j}}\mathbf{q}_{\mathbf{\Omega}^{j}} = \mathbf{\Phi}\mathbf{x} + \mathbf{w} - \mathbf{\Phi}_{\mathbf{\Omega}^{j}}\mathbf{q}_{\mathbf{\Omega}^{j}}$. It then follows

$$\begin{aligned}\left\|\mathbf{r}^{j-1}\right\|_{2} &= \left\|\mathbf{\Phi}\mathbf{x} + \mathbf{w} - \mathbf{\Phi}_{\mathbf{\Omega}^{j-1}}\mathbf{q}_{\mathbf{\Omega}^{j-1}}\right\|_{2} = \left\|\mathbf{\Phi}_{\mathbf{\Omega}^{j-1}}\mathbf{x}_{\mathbf{\Omega}^{j-1}} + \mathbf{\Phi}_{T\backslash\mathbf{\Omega}^{j-1}}\mathbf{x}_{T\backslash\mathbf{\Omega}^{j-1}} + \mathbf{w} - \mathbf{\Phi}_{\mathbf{\Omega}^{j-1}}\mathbf{q}_{\mathbf{\Omega}^{j-1}}\right\|_{2}\\
&= \left\|\mathbf{\Phi}(\tilde{\mathbf{x}}_{\mathbf{\Omega}^{j-1}} + \tilde{\mathbf{x}}_{T\backslash\mathbf{\Omega}^{j-1}} - \tilde{\mathbf{q}}_{\mathbf{\Omega}^{j-1}}) + \mathbf{w}\right\|_{2} \geq \left\|\mathbf{\Phi}(\tilde{\mathbf{x}}_{T\backslash\mathbf{\Omega}^{j-1}} + \tilde{\mathbf{x}}_{\mathbf{\Omega}^{j-1}} - \tilde{\mathbf{q}}_{\mathbf{\Omega}^{j-1}})\right\|_{2} - \left\|\mathbf{w}\right\|_{2}\\
&\stackrel{(a)}{\geq} \sqrt{1-\delta_{3K}}\left\|\tilde{\mathbf{x}}_{T\backslash\mathbf{\Omega}^{j-1}} + \tilde{\mathbf{x}}_{\mathbf{\Omega}^{j-1}} - \tilde{\mathbf{q}}_{\mathbf{\Omega}^{j-1}}\right\|_{2} - \left\|\mathbf{w}\right\|_{2}\\
&\stackrel{(b)}{=} \sqrt{1-\delta_{3K}}\sqrt{\left\|\tilde{\mathbf{x}}_{T\backslash\mathbf{\Omega}^{j-1}}\right\|_{2}^{2} + \left\|\tilde{\mathbf{x}}_{\mathbf{\Omega}^{j-1}} - \tilde{\mathbf{q}}_{\mathbf{\Omega}^{j-1}}\right\|_{2}^{2}} - \left\|\mathbf{w}\right\|_{2}\\
&\geq \sqrt{1-\delta_{3K}}\left\|\tilde{\mathbf{x}}_{T\backslash\mathbf{\Omega}^{j-1}}\right\|_{2} - \left\|\mathbf{w}\right\|_{2}, \end{aligned} \tag{A.104}$$

where (a) follows since $\mathbf{\Phi}$ satisfies RIP of order $3K$ and (b) holds because $\tilde{\mathbf{x}}_{T\backslash\mathbf{\Omega}^{j-1}}$ and $(\tilde{\mathbf{x}}_{\mathbf{\Omega}^{j-1}} - \tilde{\mathbf{q}}_{\mathbf{\Omega}^{j-1}})$ have disjoint supports. We claim that

$$\left\|\mathbf{x}_{T\backslash\mathbf{\Omega}^{j}}\right\|_{2} \leq \alpha \left\|\mathbf{x}_{T\backslash\mathbf{\Omega}^{j-1}}\right\|_{2} + \beta \left\|\mathbf{w}\right\|_{2}, \tag{A.105}$$

where $\alpha$ and $\beta$ are defined in, respectively, (4.17) and (4.18). It then follows from (A.103) and (A.105) that

$$\left\|\mathbf{r}^{j}\right\|_{2} \leq \sqrt{1+\delta_{3K}}\alpha\left\|\mathbf{x}_{T\backslash\mathbf{\Omega}^{j-1}}\right\|_{2} + (1 + \sqrt{1+\delta_{3K}}\beta)\left\|\mathbf{w}\right\|_{2}. \tag{A.106}$$

By assumption, the SP algorithm terminates after $l$ iterations, i.e., $\left\|\mathbf{r}^{l}\right\|_{2} \geq \left\|\mathbf{r}^{l-1}\right\|_{2}$. Then, from (A.104) and (A.106), we obtain

$$(\sqrt{1-\delta_{3K}} - \sqrt{1+\delta_{3K}}\alpha)\left\|\mathbf{x}_{T\backslash\mathbf{\Omega}^{l}}\right\|_{2} \leq (2 + \sqrt{1+\delta_{3K}}\beta)\left\|\mathbf{w}\right\|_{2}. \tag{A.107}$$

Since $(\sqrt{1-\delta_{3K}} - \sqrt{1+\delta_{3K}}\alpha) > 0$ as $\delta_{3K} \leq 0.2412$ (see (A.63)), the assertion of Lemma A.7 (i.e., inequality (A.94)) directly follows from (A.107).

Now, it remains to prove the claim (A.105). The proof is basically the same as that of Lemma A.5, except that the effect of noise should be taken into account. It suffices to prove the following two conditions:

$$\left\|\mathbf{x}_{T\backslash\tilde{\mathbf{\Omega}}^{j}}\right\|_{2} \leq \frac{2\delta_{3K}}{1-\delta_{3K}}\sqrt{1+\delta_{3K}^{2}\frac{1+\delta_{3K}}{1-\delta_{3K}}}\left\|\mathbf{x}_{T\backslash\mathbf{\Omega}^{j-1}}\right\|_{2} + \frac{2\sqrt{1+\delta_{3K}}}{1-\delta_{3K}}\left\|\mathbf{w}\right\|_{2}, \tag{A.108}$$

and

$$\left\|\mathbf{x}_{T\setminus\mathbf{\Omega}^j}\right\|_2 \leq \sqrt{1 + \frac{4\delta_{3K}^2(1+\delta_{3K})}{1-\delta_{3K}}}\left\|\mathbf{x}_{T\setminus\breve{\mathbf{\Omega}}^j}\right\|_2 + \frac{2}{\sqrt{1-\delta_{3K}}}\|\mathbf{w}\|_2. \quad (A.109)$$

Inequality (A.105) follows immediately by substituting the upper bound on $\left\|\mathbf{x}_{T\setminus\breve{\mathbf{\Omega}}^j}\right\|_2$ given in (A.108) into (A.109) together with some straightforward manipulations.

*(i) Derivations of (A.108):* Let us revisit (A.77), which is a consequence of Step 3.1 of the SP algorithm (see (A.75)~(A.77)). When noise is present, an upper bound of the RHS of (A.77) can be obtained as

$$\left\|\mathbf{\Phi}^*_{\mathbf{\Omega}^j_\Delta \setminus T}\mathbf{r}^{j-1}\right\|_2 \stackrel{(a)}{=} \left\|\mathbf{\Phi}^*_{\mathbf{\Omega}^j_\Delta \setminus T}(\mathbf{I} - \mathbf{P}_{\mathbf{\Omega}^{j-1}})(\mathbf{\Phi}\mathbf{x} + \mathbf{w})\right\|_2 \leq \left\|\mathbf{\Phi}^*_{\mathbf{\Omega}^j_\Delta \setminus T}(\mathbf{I} - \mathbf{P}_{\mathbf{\Omega}^{j-1}})\mathbf{\Phi}\mathbf{x}\right\|_2 + \left\|\mathbf{\Phi}^*_{\mathbf{\Omega}^j_\Delta \setminus T}(\mathbf{I} - \mathbf{P}_{\mathbf{\Omega}^{j-1}})\mathbf{w}\right\|_2, \quad (A.110)$$

where (a) is due to (A.102). Similarly, a lower bound of the LHS of (A.77) can be obtained as

$$\left\|\mathbf{\Phi}^*_{T \setminus \mathbf{\Omega}^j_\Delta}\mathbf{r}^{j-1}\right\|_2 = \left\|\mathbf{\Phi}^*_{T \setminus \mathbf{\Omega}^j_\Delta}(\mathbf{I} - \mathbf{P}_{\mathbf{\Omega}^{j-1}})(\mathbf{\Phi}\mathbf{x} + \mathbf{w})\right\|_2 \geq \left\|\mathbf{\Phi}^*_{T \setminus \mathbf{\Omega}^j_\Delta}(\mathbf{I} - \mathbf{P}_{\mathbf{\Omega}^{j-1}})\mathbf{\Phi}\mathbf{x}\right\|_2 - \left\|\mathbf{\Phi}^*_{T \setminus \mathbf{\Omega}^j_\Delta}(\mathbf{I} - \mathbf{P}_{\mathbf{\Omega}^{j-1}})\mathbf{w}\right\|_2. \quad (A.111)$$

From (A.110), (A.111) and (A.77), we have

$$\left\|\mathbf{\Phi}^*_{T \setminus \mathbf{\Omega}^j_\Delta}(\mathbf{I} - \mathbf{P}_{\mathbf{\Omega}^{j-1}})\mathbf{\Phi}\mathbf{x}\right\|_2 \leq \left\|\mathbf{\Phi}^*_{\mathbf{\Omega}^j_\Delta \setminus T}(\mathbf{I} - \mathbf{P}_{\mathbf{\Omega}^{j-1}})\mathbf{\Phi}\mathbf{x}\right\|_2 + \left\|\mathbf{\Phi}^*_{\mathbf{\Omega}^j_\Delta \setminus T}(\mathbf{I} - \mathbf{P}_{\mathbf{\Omega}^{j-1}})\mathbf{w}\right\|_2 + \left\|\mathbf{\Phi}^*_{T \setminus \mathbf{\Omega}^j_\Delta}(\mathbf{I} - \mathbf{P}_{\mathbf{\Omega}^{j-1}})\mathbf{w}\right\|_2. \quad (A.112)$$

Note that, for $S_1 \subset \{1,...,N\}$ with $|S_1| \leq K$, we have

$$\left\|\mathbf{\Phi}^*_{S_1}(\mathbf{I} - \mathbf{P}_{\mathbf{\Omega}^{j-1}})\mathbf{w}\right\|_2 \stackrel{(a)}{\leq} \sqrt{1+\delta_{3K}}\left\|(\mathbf{I} - \mathbf{P}_{\mathbf{\Omega}^{j-1}})\mathbf{w}\right\|_2 \leq \sqrt{1+\delta_{3K}}\|\mathbf{w}\|_2, \quad (A.113)$$

where (a) follows from Lemma A.1. Hence, from (A.112) and (A.113), it follows

$$\left\|\mathbf{\Phi}^*_{T \setminus \mathbf{\Omega}^j_\Delta}(\mathbf{I} - \mathbf{P}_{\mathbf{\Omega}^{j-1}})\mathbf{\Phi}\mathbf{x}\right\|_2 \leq \left\|\mathbf{\Phi}^*_{\mathbf{\Omega}^j_\Delta \setminus T}(\mathbf{I} - \mathbf{P}_{\mathbf{\Omega}^{j-1}})\mathbf{\Phi}\mathbf{x}\right\|_2 + 2\sqrt{1+\delta_{3K}}\|\mathbf{w}\|_2. \quad (A.114)$$

Since $(\mathbf{I} - \mathbf{P}_{\mathbf{\Omega}^{j-1}})\mathbf{\Phi}\mathbf{x}$ is the residual $\mathbf{r}^{j-1}$ in the noiseless case (see (A.59)), based on (A.114) and by following essentially the same procedures as (A.78)~(A.80) we then obtain (A.108).

*(ii) Derivations of (A.109):*

The proof procedures are similar to the derivation of (A.68). We first revisit (A.82), which is a consequence of Step 3.5 of the SP algorithm (see (A.81)~(A.82)). An upper bound for the RHS of (A.82) can be derived as follows

$$\left\|\breve{\mathbf{q}}_{\breve{\mathbf{\Omega}}^j \setminus T}\right\|_2 \leq \frac{\delta_{3K}\sqrt{1+\delta_{3K}}}{\sqrt{1-\delta_{3K}}}\left\|\mathbf{x}_{T\setminus\breve{\mathbf{\Omega}}^j}\right\|_2 + \frac{1}{\sqrt{1-\delta_{3K}}}\|\mathbf{w}\|_2; \quad (A.115)$$

also, a lower bound for the LHS of (A.82) can be obtained as

$$\left\| \breve{\mathbf{q}}_{\bar{\Omega}^j} \right\|_2 \geq \sqrt{\left\| \mathbf{x}_{T\setminus\Omega^j} \right\|_2^2 - \left\| \mathbf{x}_{T\setminus\breve{\Omega}^j} \right\|_2^2} - \frac{\delta_{3K}\sqrt{1+\delta_{3K}}}{\sqrt{1-\delta_{3K}}} \left\| \mathbf{x}_{T\setminus\breve{\Omega}^j} \right\|_2 - \frac{1}{\sqrt{1-\delta_{3K}}} \|\mathbf{w}\|_2. \quad \text{(A.116)}$$

Then, (A.109) follows from (A.82), (A.115), and (A.116).

To show (A.115) is true, we observe that

$$\mathbf{P}_{\bar{\Omega}^j}\mathbf{y} = \mathbf{P}_{\bar{\Omega}^j}(\mathbf{\Phi}\mathbf{x} + \mathbf{w}) = \mathbf{P}_{\bar{\Omega}^j}(\mathbf{\Phi}_{\bar{\Omega}^j}\mathbf{x}_{\bar{\Omega}^j} + \mathbf{\Phi}_{T\setminus\bar{\Omega}^j}\mathbf{x}_{T\setminus\bar{\Omega}^j} + \mathbf{w}) = \mathbf{\Phi}_{\bar{\Omega}^j}\mathbf{x}_{\bar{\Omega}^j} + \mathbf{P}_{\bar{\Omega}^j}\mathbf{\Phi}_{T\setminus\bar{\Omega}^j}\mathbf{x}_{T\setminus\bar{\Omega}^j} + \mathbf{P}_{\bar{\Omega}^j}\mathbf{w}. \quad \text{(A.117)}$$

Note that there exist $\mathbf{z} \in \mathbb{R}^{2K}$ and $\mathbf{u} \in \mathbb{R}^{2K}$ such that

$$\mathbf{P}_{\bar{\Omega}^j}\mathbf{\Phi}_{T\setminus\bar{\Omega}^j}\mathbf{x}_{T\setminus\bar{\Omega}^j} = \mathbf{\Phi}_{\bar{\Omega}^j}\mathbf{z} = \mathbf{\Phi}\tilde{\mathbf{z}} \quad \text{and} \quad \mathbf{P}_{\bar{\Omega}^j}\mathbf{w} = \mathbf{\Phi}_{\bar{\Omega}^j}\mathbf{u} = \mathbf{\Phi}\tilde{\mathbf{u}}, \quad \text{(A.118)}$$

where $\tilde{\mathbf{z}}$ and $\tilde{\mathbf{u}}$ are obtained by padding zeros to, respectively, $\tilde{\mathbf{z}}$ and $\tilde{\mathbf{u}}$. With (A.118), we can then rewrite (A.117) as

$$\mathbf{P}_{\bar{\Omega}^j}\mathbf{y} = \mathbf{\Phi}_{\bar{\Omega}^j}(\mathbf{x}_{\bar{\Omega}^j} + \mathbf{z} + \mathbf{u}) = \mathbf{\Phi}\underbrace{(\tilde{\mathbf{x}}_{\bar{\Omega}^j} + \tilde{\mathbf{z}} + \tilde{\mathbf{u}})}_{\breve{\mathbf{q}} \text{ in Step 3.5}}. \quad \text{(A.119)}$$

Then, we obtain

$$\left\| \breve{\mathbf{q}}_{\bar{\Omega}^j\setminus T} \right\|_2 \stackrel{(a)}{=} \left\| (\tilde{\mathbf{z}} + \tilde{\mathbf{u}} + \tilde{\mathbf{x}}_{\bar{\Omega}^j})_{\bar{\Omega}^j\setminus T} \right\|_2 \stackrel{(b)}{=} \left\| (\tilde{\mathbf{z}} + \tilde{\mathbf{u}})_{\bar{\Omega}^j\setminus T} \right\|_2 \leq \left\| \tilde{\mathbf{z}} + \tilde{\mathbf{u}} \right\|_2, \quad \text{(A.120)}$$

where (a) follows from (A.119) and (b) holds since only those entries of $\mathbf{x}$ indexed by $T$ are nonzero. Also, we have

$$\sqrt{1-\delta_{3K}} \|\tilde{\mathbf{z}}\|_2 \stackrel{(a)}{\leq} \|\mathbf{\Phi}\tilde{\mathbf{z}}\|_2 \stackrel{(b)}{=} \left\| \mathbf{P}_{\bar{\Omega}^j}\mathbf{\Phi}_{T\setminus\bar{\Omega}^j}\mathbf{x}_{T\setminus\bar{\Omega}^j} \right\|_2 = \left\| \mathbf{P}_{\bar{\Omega}^j}\mathbf{\Phi}\tilde{\mathbf{x}}_{T\setminus\bar{\Omega}^j} \right\|_2$$

$$\stackrel{(c)}{\leq} \delta_{3K}\sqrt{1+\delta_{3K}} \left\| \tilde{\mathbf{x}}_{T\setminus\bar{\Omega}^j} \right\|_2 = \delta_{3K}\sqrt{1+\delta_{3K}} \left\| \mathbf{x}_{T\setminus\bar{\Omega}^j} \right\|_2, \quad \text{(A.121)}$$

where (a) follows from RIP, (b) holds from (A.118), and (c) is due to Lemma A.4. Also, since $\mathbf{\Phi}$ satisfies RIP, it follows from (A.118) that

$$\sqrt{1-\delta_{3K}} \|\tilde{\mathbf{u}}\|_2 \leq \|\mathbf{\Phi}\tilde{\mathbf{u}}\|_2 = \left\| \mathbf{P}_{\bar{\Omega}^j}\mathbf{w} \right\|_2 \leq \|\mathbf{w}\|_2. \quad \text{(A.122)}$$

With (A.121) and (A.122), it follows

$$\|\tilde{\mathbf{z}} + \tilde{\mathbf{u}}\|_2 \leq \|\tilde{\mathbf{z}}\|_2 + \|\tilde{\mathbf{u}}\|_2 \stackrel{(a)}{\leq} \frac{\delta_{3K}\sqrt{1+\delta_{3K}}}{\sqrt{1-\delta_{3K}}} \left\| \mathbf{x}_{T\setminus\bar{\Omega}^j} \right\|_2 + \frac{1}{\sqrt{1-\delta_{3K}}} \|\mathbf{w}\|_2. \quad \text{(A.123)}$$

The assertion (A.115) follows from (A.120) and (A.123). To prove (A.116), we first observe that

$$\begin{aligned}
\left\|\breve{\mathbf{q}}_{\bar{\mathbf{\Omega}}^j}\right\|_2 &\stackrel{(a)}{=} \left\|(\tilde{\mathbf{z}} + \tilde{\mathbf{u}} + \tilde{\mathbf{x}}_{T\cap\bar{\mathbf{\Omega}}^j})_{\bar{\mathbf{\Omega}}^j}\right\|_2 \\
&\geq \left\|(\tilde{\mathbf{x}}_{T\cap\bar{\mathbf{\Omega}}^j})_{\bar{\mathbf{\Omega}}^j}\right\|_2 - \left\|(\tilde{\mathbf{z}} + \tilde{\mathbf{u}})_{\bar{\mathbf{\Omega}}^j}\right\|_2 \geq \left\|(\tilde{\mathbf{x}}_{T\cap\bar{\mathbf{\Omega}}^j})_{\bar{\mathbf{\Omega}}^j}\right\|_2 - \left\|\tilde{\mathbf{z}} + \tilde{\mathbf{u}}\right\|_2 \\
&\stackrel{(b)}{=} \left\|(\tilde{\mathbf{x}}_{\bar{\mathbf{\Omega}}^j})_{\bar{\mathbf{\Omega}}^j}\right\|_2 - \left\|\tilde{\mathbf{z}} + \tilde{\mathbf{u}}\right\|_2 \\
&\stackrel{(c)}{=} \left\|\tilde{\mathbf{x}}_{\bar{\mathbf{\Omega}}^j}\right\|_2 - \left\|\tilde{\mathbf{z}} + \tilde{\mathbf{u}}\right\|_2,
\end{aligned} \tag{A.124}$$

where (a) follows from (A.119), (b) is true since $\mathbf{x}$ is supported on $T$, and (c) follows since $\bar{\mathbf{\Omega}}^j = \breve{\mathbf{\Omega}}^j \setminus \mathbf{\Omega}^j \subset \breve{\mathbf{\Omega}}^j$. With the aid of (A.93), (A.123), and (A.124), we have (A.116). □